\documentclass[twocolumn,showpacs,preprintnumbers,superscriptaddress,amsmath,amssymb]{revtex4}

\usepackage{graphicx}
\usepackage{dcolumn}
\usepackage{bm}
\usepackage{graphicx,epstopdf}
\usepackage{xcolor}
\usepackage{textcomp}
\usepackage{soul}
\usepackage{csquotes}
\usepackage{mathrsfs} 
\usepackage{amsmath}
\usepackage{bigints} 
\usepackage{amsthm,amssymb} 
\usepackage[utf8]{inputenc}
\usepackage[english]{babel}
\usepackage[shortlabels]{enumitem}
\usepackage{mathrsfs,bigints,mathtools,graphicx}
\newcommand\norm[1]{\left\lVert#1\right\rVert}

\begin{document}


\title{Transitions from chimeras to coherence: An analytical approach by means of the coherent stability function}

\author{Sarbendu Rakshit}
\affiliation{Physics and Applied Mathematics Unit, Indian Statistical Institute, 203 B. T. Road, Kolkata 700108, India}

\author{Zahra Faghani}
\affiliation{Department of Biomedical Engineering, Amirkabir University of Technology, 424 Hafez Ave., Tehran 15875-4413, Iran}

\author{Fatemeh Parastesh}
\affiliation{Department of Biomedical Engineering, Amirkabir University of Technology, 424 Hafez Ave., Tehran 15875-4413, Iran}

\author{Shirin Panahi}
\affiliation{Department of Biomedical Engineering, Amirkabir University of Technology, 424 Hafez Ave., Tehran 15875-4413, Iran}

\author{Sajad Jafari}
\affiliation{Department of Biomedical Engineering, Amirkabir University of Technology, 424 Hafez Ave., Tehran 15875-4413, Iran}

\author{Dibakar Ghosh}\email{dibakar@isical.ac.in}
\affiliation{Physics and Applied Mathematics Unit, Indian Statistical Institute, 203 B. T. Road, Kolkata 700108, India}

\author{Matja{\v z} Perc}
\email{matjaz.perc@um.si}
\affiliation{Faculty of Natural Sciences and Mathematics, University of Maribor, Koro{\v s}ka cesta 160, SI-2000 Maribor, Slovenia}
\affiliation{Complexity Science Hub Vienna, Josefst{\"a}dterstra{\ss}e 39, A-1080 Vienna, Austria}
\date{\today}

\begin{abstract}
Chimera states have been a vibrant subject of research in the recent past, but the analytical treatment of transitions from chimeras to coherent states remains a challenge. Here we analytically derive the necessary conditions for this transition by means of the coherent stability function approach, which is akin to the master stability function approach that is traditionally used to study the stability of synchronization in coupled oscillators. In chimera states, there is typically at least one group of oscillators that evolves in a drifting, random manner, while other groups of oscillators follow a smoother, more coherent profile. In the coherent state, there thus exists a smooth functional relationship between the oscillators. This lays the foundation for the coherent stability function approach, where we determine the stability of the coherent state. We subsequently test the analytical prediction numerically by calculating the strength of incoherence during the transition point. We use leech neurons, which exhibit a coexistence of chaotic and periodic tonic spiking depending on initial conditions, coupled via non-local electrical synapses, to demonstrate our approach. We systematically explore various dynamical states with the focus on the transitions between chimeras and coherence, fully confirming the validity of the coherent stability function. We also observe complete synchronization for higher values of the coupling strength, which we verify by the master stability function approach.
\end{abstract}

\maketitle

\section{Introduction}
\par Chimera states are emergent spatiotemporal patterns consisting of the coexistence of coherent and drifting domains in a network of identical oscillators. They were first observed in non-locally coupled identical phase oscillators by Kuramoto and Battogtokh \cite{r12}.  Later these patterns was named chimera states by Abrams and Strogatz \cite{r13}. With the invention of these states, research on the emergent behaviors of a network of coupled oscillators has been reanimated. Chimera states have been studied extensively in recent years \cite{r15a,chimera_r1,r14a,chimera_r2}. They have been investigated in different types of oscillators, such as phase oscillators \cite{r18}, chaotic and periodic maps \cite{r19}, pendulum oscillators \cite{r20}, neuronal models \cite{r15a}, mechanical systems \cite{r21}, and chemical systems \cite{r22}. Since their discovery, several types of chimera states have been identified based on different schemes of spatio-temporal patterns. An amplitude chimera state \cite{amplt_cmra} is one in which in the incoherent subpopulation, the amplitudes of the oscillators differ, but the phase velocity of the entire network remains the same. A spiral wave chimera \cite{r31,epjst} is a two-dimensional chimera consisting of a randomizing desynchronized core surrounded by phase-locked oscillators. Chimera death is a partially coherent inhomogeneous pattern consisting of spatially coherent and incoherent oscillation death \cite{r25}. In a stationary chimera state, the position of coherent and incoherent clusters is fixed. However, in a non-stationary chimera state, the position of coherent and incoherent clusters changes over time \cite{r40}. There are other types of non-stationary chimera states, such as breathing chimera \cite{r18}, imperfect chimera \cite{r20}, traveling chimera \cite{r46}, multi-headed chimera \cite{r29} and alternating chimera \cite{r30}. Recently, a new type of nonstationary chimera state called a spike chimera state \cite{spike} has also been discovered, where the chimera and fully coherent states appear alternately with respect to time in a neuronal hypernetwork. In the context of chimera states, the underlying network architecture and the coupling function also play an important role. With linear diffusive coupling, the chimera states emerge only for non-local \cite{r13} or global \cite{chimera_global} network architecture, while for non-linear chemical synaptic coupling, chimera states can emerge irrespective of local, non-local, global \cite{r39}, and even unidirectional local and non-local coupling \cite{spike}. Chimera states have also has been observed in many real biological and physical systems, including ecology \cite{cmra_eco1,r29}, SQUID metamaterial \cite{cmra_SQUID}, and quantum systems \cite{cmra_quantum}.

\par The nervous system is a complex biological process that consists of billions of neurons. Neurons as systems with strong non-linear dynamical properties can generate various complex activities from a rest state to tonic spiking and bursting states \cite{r7,r9}. Among these non-linear complex activities, a ubiquitous observed pattern is tonic spiking, which correlates to different functions of the nervous systems, such as the coding of sensory information, information processing, memory formation, attention, and motor control \cite{r8}. There are many mathematical models that effort to illustrate and investigate different behaviors of neurons \cite{r1}. One effective way to gain a better understanding of the behaviors of the entire nervous system is to investigate them in the network of coupled neurons \cite{r2}. Such investigations, which provide the advantage of considering the interaction among all particles of the network, are significant and useful tools to perceive the potential mechanism of neuronal diseases \cite{r6}. Several mathematical neuronal models can exhibit the coexistence of tonic spiking, bursting, and rest states, which has been studied extensively \cite{r2}. In a dynamical system, the coexistence of several attractors in phase space is referred to as multi-stability, where each attractor is determined by initial conditions. Lencher et al. \cite{r10} exhibited bistability in a Leech neuron via experiments. The isolated Leech neuronal model exhibits bistable states, consisting of one chaotic and one periodic states, depending on the initial conditions \cite{r7}.

\par Previous studies have shown that a chimera state is closely related to various types of brain phenomena, such as hemisphere sleep in animals, epilepsy, Parkinson's, Alzheimer's, and schizophrenia \cite{r32,r33,r34}. Chimera states also play an important role in neuroscience and cardiac fibrillation \cite{r35}. Andrzejak et al. \cite{r33} exhibited an abrupt conversion of the coexistence of coherent and incoherent oscillators into completely coherent oscillators in an epileptic seizure in humans. It has been shown that spiral wave chimera states appear in the contractions of heart cells during ventricular fibrillation, which is formed on the surface of a sphere  \cite{chimera_r1}. Chimera states have also been observed in unihemispheric slow-wave sleep of some aquatic mammals and a few migratory birds \cite{cmra_unih1,cmra_unih2}. During slow-wave sleep, one-half of the brain is awake and the neurons in that portion are desynchronized, while the other half is at rest and the corresponding neurons are synchronized. Chimera states have also been investigated in neuronal networks, such as the Hindmarsh-Rose \cite{r37,r37a,r39} and FitzHugh-Nagumo network \cite{r38,r38a}. V\"{u}llings et al. \cite{r37} considered a two-dimensional network of Hindmarsh-Rose neurons by a rotating coupling matrix, representatives of excitability type I, and they obtained different multi-headed chimera patterns. Omelchenko et al. \cite{r38} constructed a network of identical non-locally coupled Fitzhugh-Nagumo oscillators and found new types of chimera states.

\par Motivated by the above, in this paper, we investigate chimera states in an identical Leech neuronal network that interact through non-local electrical synapses. The non-local coupling emerges in several real-world applications ranging from neuronal networks \cite{neuronal_nl}, more particularly in ocular dominance strips \cite{ocular_nl}, to chemical oscillators \cite{r12} and Josephson function arrays \cite{jj_nl}. In particular, the appearance of non-stationary chimera patterns is observed, which have an important role in neural networks and need more attention \cite{r41}. The most interesting part of our work is the analysis of the instability of chimera states towards coherent states.

\par In our observed non-stationary chimera states, the entire network splits into two groups, one with an incoherent domain and the other with a coherent domain. Contrary to these chimera states, all the oscillators in the coherent state follow a smooth coherent profile, i.e., they follow an amplitude-locked oscillation. From this, we claim that there may exist a smooth functional relationship among the oscillators. That is, the network of oscillators follows the generalized synchronization (GS) state. This is a typical coordination of the rhythms among the coupled oscillators. This type of synchronization was first noticed by Rulkov et al. \cite{ausys_3} between two unidirectionally coupled systems.  The auxiliary system approach \cite{ausys_1} has been widely used to detect and analyze the stability of GS states. At first, this approach was proposed in a drive-response system \cite{ausys_1}. The necessary and sufficient conditions for GS based on the auxiliary system approach were later illustrated in Ref. \cite{ausys_4}.

\par Later, this seminal approach was extended to bidirectionally coupled systems and complex networks by introducing a system consisting of an auxiliary node for each original node \cite{ausys_2}. In Ref. \cite{hung}, the authors uncovered GS in a bidirectional scale-free network of identical chaotic oscillators using an auxiliary-system approach. They noticed that the GS starts from the hubs of the network and then spreads throughout the whole network upon increasing the coupling strength. The occurrence and development of GS on various mutually coupled complex networks, including scale-free networks, small-world networks, random networks, and modular networks are also investigated \cite{guan}. It has been shown that, this phenomenon generally takes place in such networks for both coupled identical and nonidentical oscillators. For coupled identical oscillators, they found a typical path from non-synchronization to complete synchronization (CS) through GS, i.e. nonsynchronization $\rightarrow$ GS $\rightarrow$ CS. In Ref. \cite{chen}, the authors investigated GS in two typical complex dynamical networks, namely small-world networks and scale-free networks, in terms of an impulsive control strategy, and they analyzed how the development of GS depends on impulsive control gains. By applying the auxiliary-system approach, they demonstrated their theoretical analysis. GS was also investigated via the adaptive bidirectional couplings strategy \cite{liu}. Adopting the auxiliary-system approach and the Lyapunov function method, the authors proved that for any given initial coupling strengths, GS can take place in complex networks consisting of nonidentical dynamical systems. Subsequently, this state was further characterized by analytical and numerical experiments in mutually coupled networks \cite{gs_njp}. Recently, Moskalenko et al.\cite{olga} compared the applicability of the auxiliary-system approach in unidirectional and bidirectional coupled systems.

\par Hence, we analyze the stability of the coherent state in terms of GS by the auxiliary system approach. The instability of chimera states indicates the stability of coherent states, while the stability of coherent states implies the instability of chimera states. Thus, we could identify the critical transition point of the chimera to the coherent state, analytically obtained by the coherent stability function approach. The effect of initial conditions is also studied by the calculation of basin stability measurement. For a higher value of the coupling strength, complete synchronization occurs in the non-locally coupled neuronal network, and it is verified by the master stability function approach.

\section{Mathematical model}\label{nonlocal}
Generally, a network of $N$ identical coupled nodes can be written as
\begin{equation}
\begin{array}{l}\label{eq_1}
\dot{\bf x}_i=F({\bf x}_i)+\dfrac{\epsilon}{d(i)}\sum\limits_{j=1}^{N}\mathscr{A}_{ij}\Gamma({\bf x}_j-{\bf x}_i),
\end{array} 
\end{equation}
where $i=1,2,\dots,N$; ${\bf x}_i$ is the state variable of node $i$. Here each isolated node modeled by a $d-$dimensional dynamical system $\dot{\bf x}=F({\bf x})$, which is a continuous vector valued function. $\epsilon$ is the coupling strength, which modulates the amount of the information propagation among the nodes, $d(i)$ is the degree of node $i$, and $\Gamma$ is the inner coupling matrix that determines which node's variable is used in the coupling. $\mathscr{A}$ is the adjacency matrix of the network. It represents the topology of the connectivity in the network, where $\mathscr{A}_{ij}=1$ if the $i^{th}$ node and the $j^{th}$ node are connected and zero otherwise. If $\mathscr{L}$ is the corresponding zero row sum Laplacian matrix of the network, then $\mathscr{L}_{ij}=-\mathscr{A}_{ij}$ if $i\ne j$ and $\mathscr{L}_{ii}=\sum_{j=1}^{N}\mathscr{A}_{ij}$. The chimera state is the broken of symmetry among the symmetrically coupled identical oscillators; only their initial conditions are different. Therefore, the underlying network is considered to be a regular network that is either local or non-local, or global. For a non-locally coupled network, the Laplacian matrix can be written as
\begin{equation}\label{lap1}
\begin{array}{lll}
	\mathscr{L}_{ij} &=-1 \hspace{10pt}{\text{for}~} 0<|i-j|\le P,\\
	&=2P\hspace{10pt}{\text{for}~} i=j,\\
	&=0\hspace{18pt}{\text{otherwise}}.
\end{array}
\end{equation}
Therefore, the degree of the $i^{th}$ node is $d(i)=2P$. For $P=1$, it is a locally coupled network and for $1<P<\frac{N}{2}$, it is non-local. The Laplacian matrix $\mathscr{L}$ is a real symmetric matrix, so all it's eigenvalues are real, and it is orthogonally diagonalizable by the basis of eigenvectors $U$. Moreover, $\mathscr{L}$ is a circulant matrix. It's $N$ eigenvalues are $\sum_{l=1}^{N}\mathscr{L}_{1l}\omega_j^{l-1}$, where $\omega_j=\exp\Big(\frac{2\pi j\sqrt{-1}}{N}\Big)$ and $j=0,1,\dots,N-1$, which yields all $N$ eigenvalues as $4\sum_{l=1}^{P}\sin^2\big(\frac{\pi l j}{N}\big)$, where $j=0,1,\dots,N-1$. For $j=0$, we get it's unique zero eigenvalues, and others are non-zeroes.

\section{Leech neuronal network}\label{math_mod}

\par A Leech neuron exhibits bi-stability of chaotic and periodic attractors. It can exhibit different neuronal activities, such as tonic spiking and bursting. Cymbalyuk et al. \cite{r7} simplified the Leech neuron in such a way that it depends on the dynamics of the potassium current $I_{K2}$ and the sodium current $I_{Na}$. This model is described by a three-dimensional differential equation
\begin{equation}
\begin{array}{l}
\label{eq.1}
\dot{V}=-\frac{1}{C}\big[\bar{g}_{K2}m^2_{K2}(V-E_K )+g_1 (V-E_1)+\\[2pt]~~~~~~~~~~~~\bar{g}_{Na}\big(f(A_1,B_1,V)\big)^3 h_{Na} (V-E_{Na} )\big],\\[5pt]
\dot{m}_{K2}=\frac{f(A_2,B_2+V^{shift}_{K2},V)-m_{K2}}{\tau_{K2}} ,\\[5pt]
\dot{h}_{Na}=\frac{f(A_3,B_3,V)-h_{Na}}{\tau_{Na}},
\end{array}
\end{equation}
where the variables $V$, $m_{K2}$, and $h_{Na}$ describe the membrane potential, the activation of $I_{K2}$ and the inactivation of $I_{Na}$ respectively. The parameters: $\bar{g}_{K2}$,  $\bar{g}_{Na}$,  and $g_1$ demonstrate the maximum conductance of $I_{K2}$, $I_{Na}$ and the conductance of the leak current, respectively. $E_k$, $E_{Na}$,  and $E_1$ refer to the reversal potential of $K^+$, $Na^+$, and the leak current, respectively. $C$ is the membrane capacitance, $\tau_{K2}$ and $\tau_{Na}$ illustrate the time constants of activation of $I_{K2} $ and inactivation of $I_{Na}$, respectively and $V^{shift}_{K2}$ characterizes the shift of the membrane potential of half-activation of $I_{K2}$ from its canonical value. $f(A,B,C)$ is the Boltzmann function \cite{r7},
\begin{equation}
\begin{array}{l}
\label{eq.2}
f(A,B,V)=\dfrac{1}{1+e^{A(B+V)}}.\\
\end{array}
\end{equation}
The values of the constant parameters that are used in this paper are represented in Table \ref{table_1}.
\begin{table}
\center
	\caption{Parameter values for the Leech neuron model (\ref{eq.1}) \cite{r7}.}
	\begin{tabular}{ccccccc}
\hline\\
Parameter  &  Value &  ~~Parameter  &  value \\
\hline\\
$\bar{g}_{K2}$ & $30~ns$ & ~~$E_1$ & $-0.046~V$ \\
\\
$\bar{g}_{Na}$ & $200~ns$ & ~~$C$ & $0.5~nF$ \\
\\
$g_{1}$ & $8~ns$ & ~~$\tau_{K2}$ & $0.25~sec$ \\
\\
$E_{K}$ & $-0.07~V$ & ~~$\tau_{Na}$ & $0.0405~sec$ \\
\\
$E_{Na}$ & $0.045~V$ & ~~$V^{shift}_{K2}$ & $-0.025361~V$ \\
\\
$A_1$ & $-150$ & ~~$B_1$ & $0.0305$ \\
\\
$A_2$ & $-83$ & ~~$B_2$ & $0.018$ \\
\\
$A_3$ & $500$ & ~~$B_3$ & $0.0333$ \\ \\
 \hline
\end{tabular}
\label{table_1}
\end{table}

\begin{figure}[b]
\centerline{\includegraphics[scale=0.43]{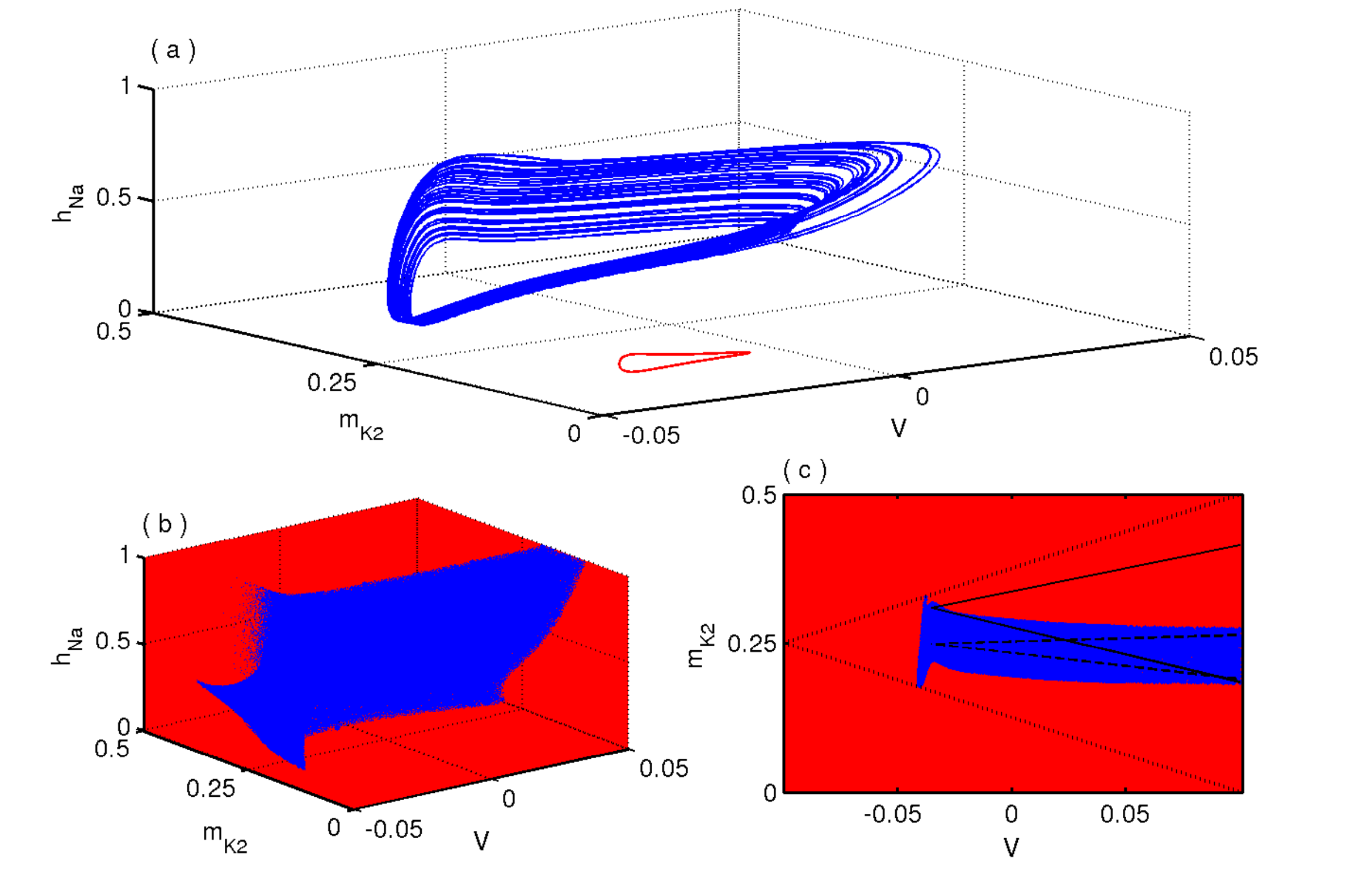}}
\caption{(Color online) (a) Three-dimensional phase space of the bistable states: chaotic attractor in blue (black) and periodic orbit in red (gray). (b) Well-separated basins of attraction of the chaotic attractor [blue (black) region] and the periodic orbit [red (gray) region]. (c) The two-dimensional basin of attraction on the $h_{Na}=0.5$ plane.  The dotted, dashed, and solid black lines, respectively, represent the V-shaped PIs, CIs and MIs.}
\label{figure_0}
\end{figure}

\begin{figure*}	 \centerline{\includegraphics[scale=0.27]{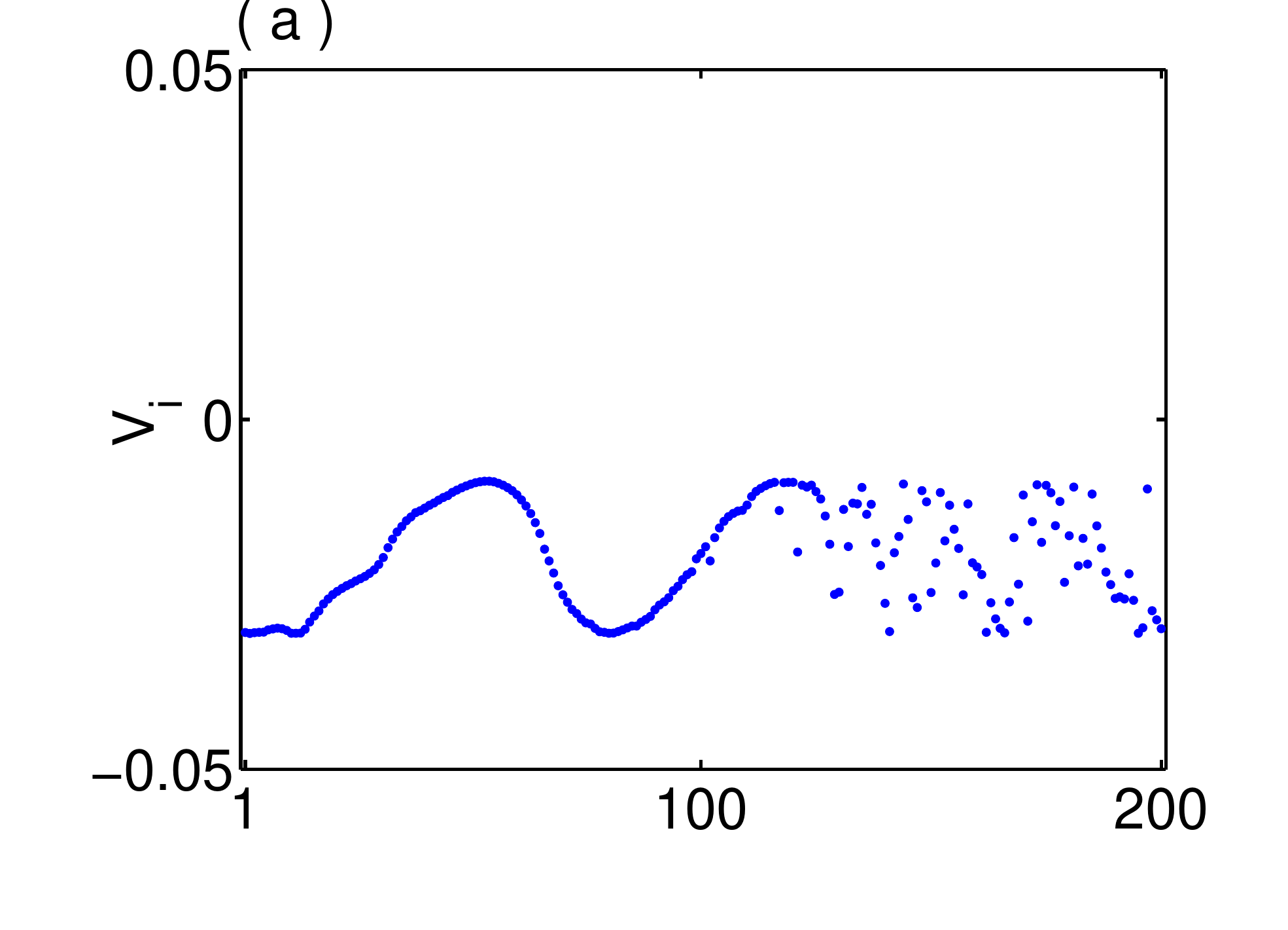}\includegraphics[scale=0.27]{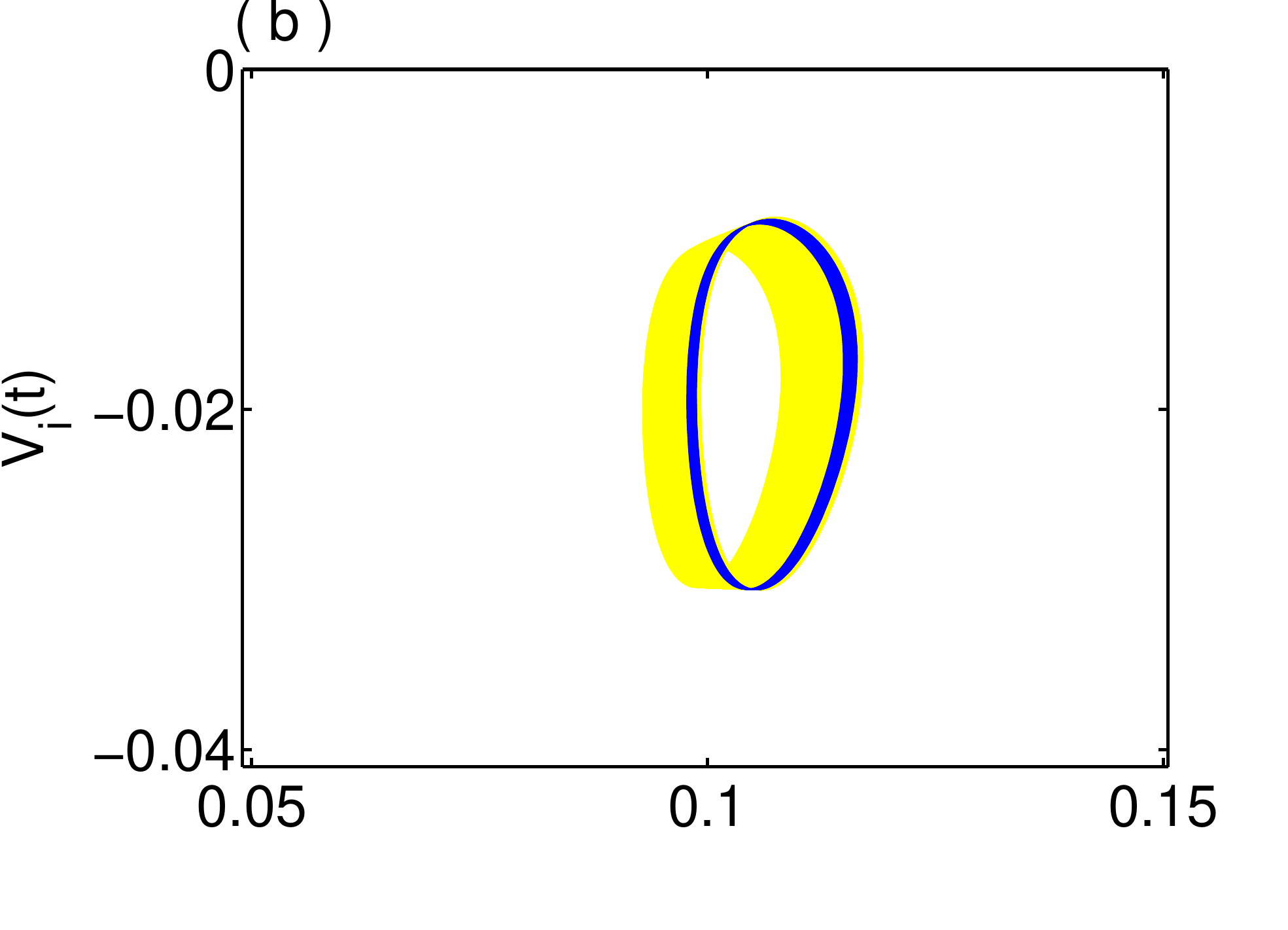}\includegraphics[scale=0.27]{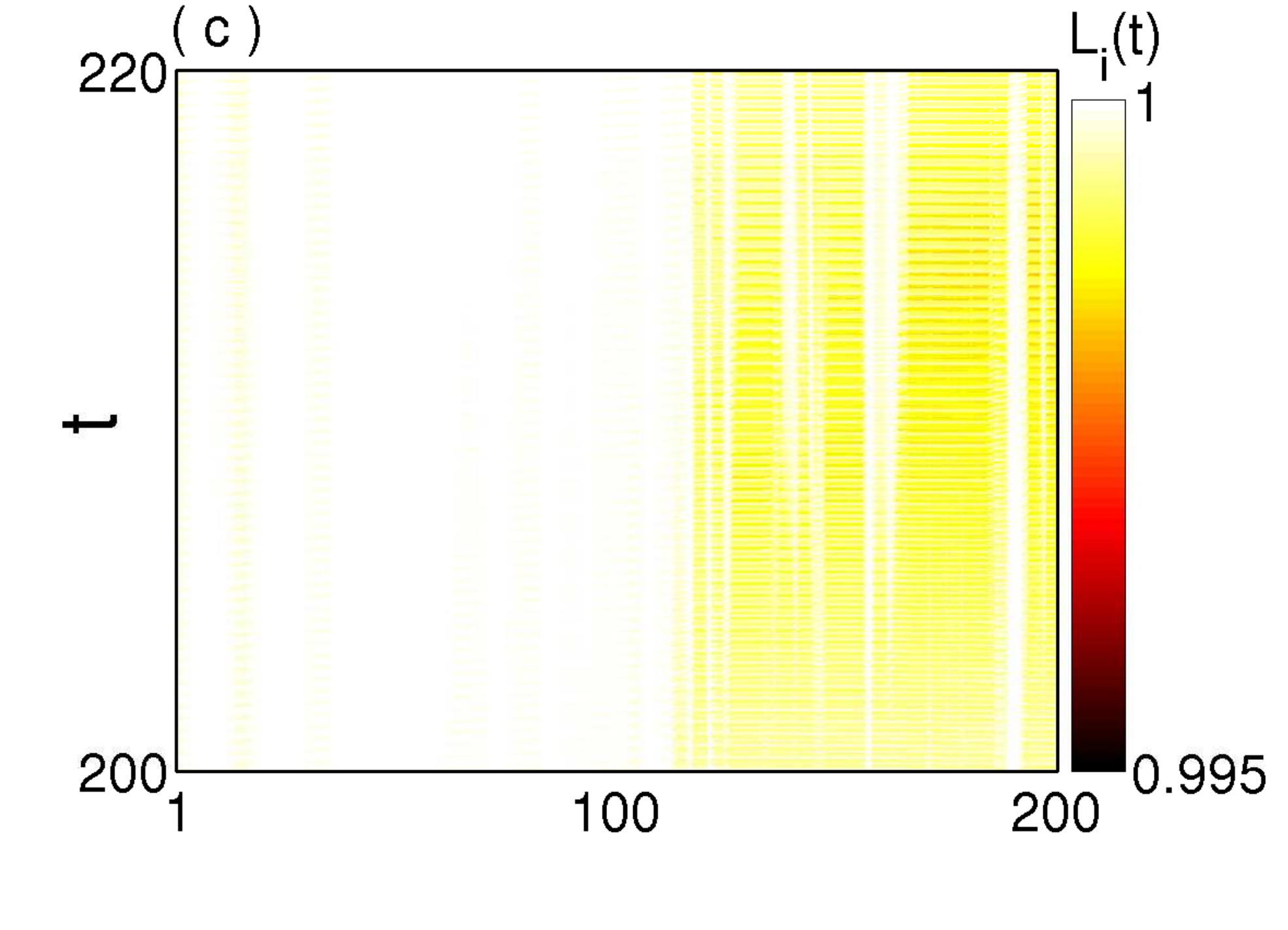}}
\centerline{\includegraphics[scale=0.27]{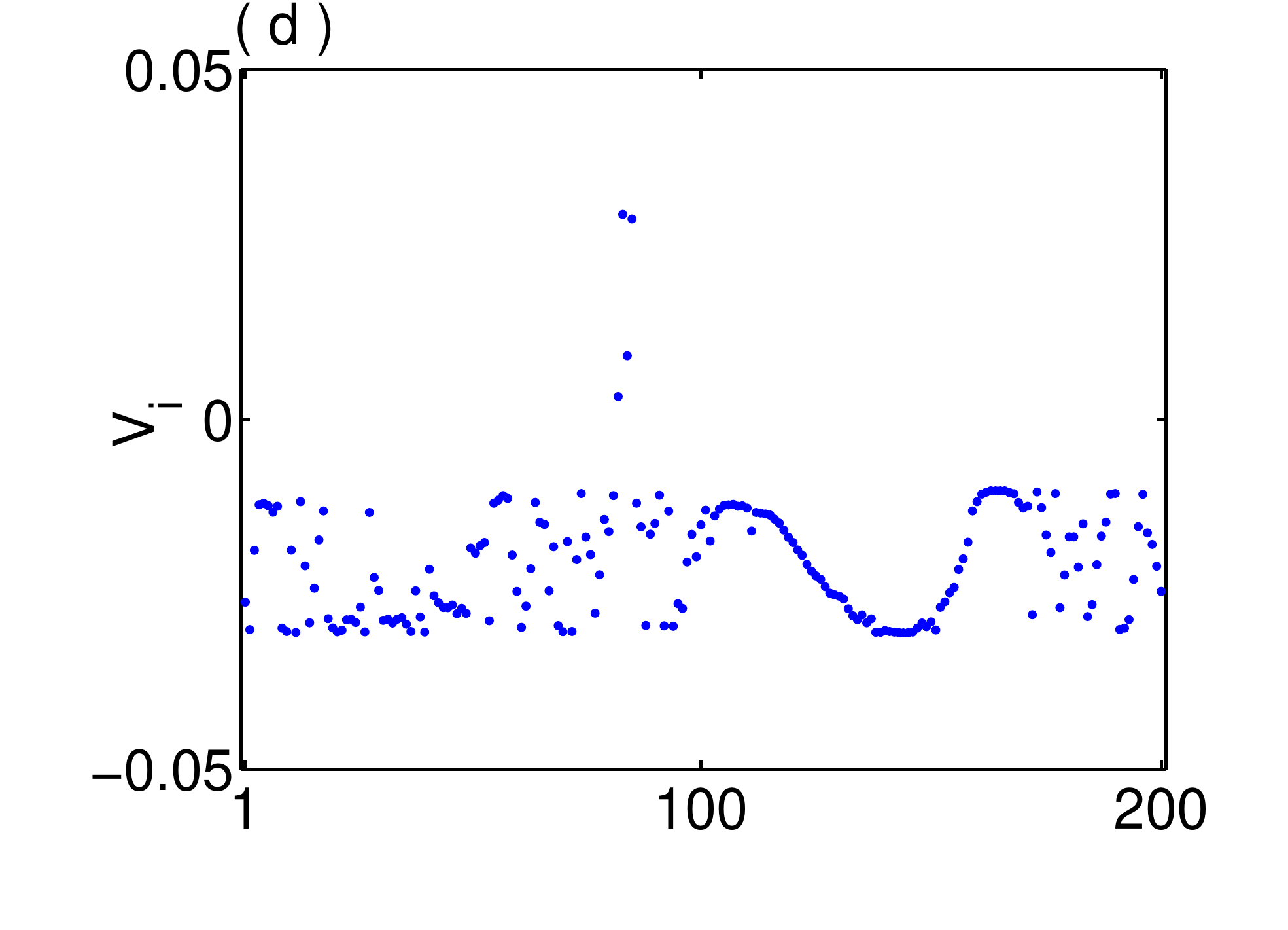}\includegraphics[scale=0.27]{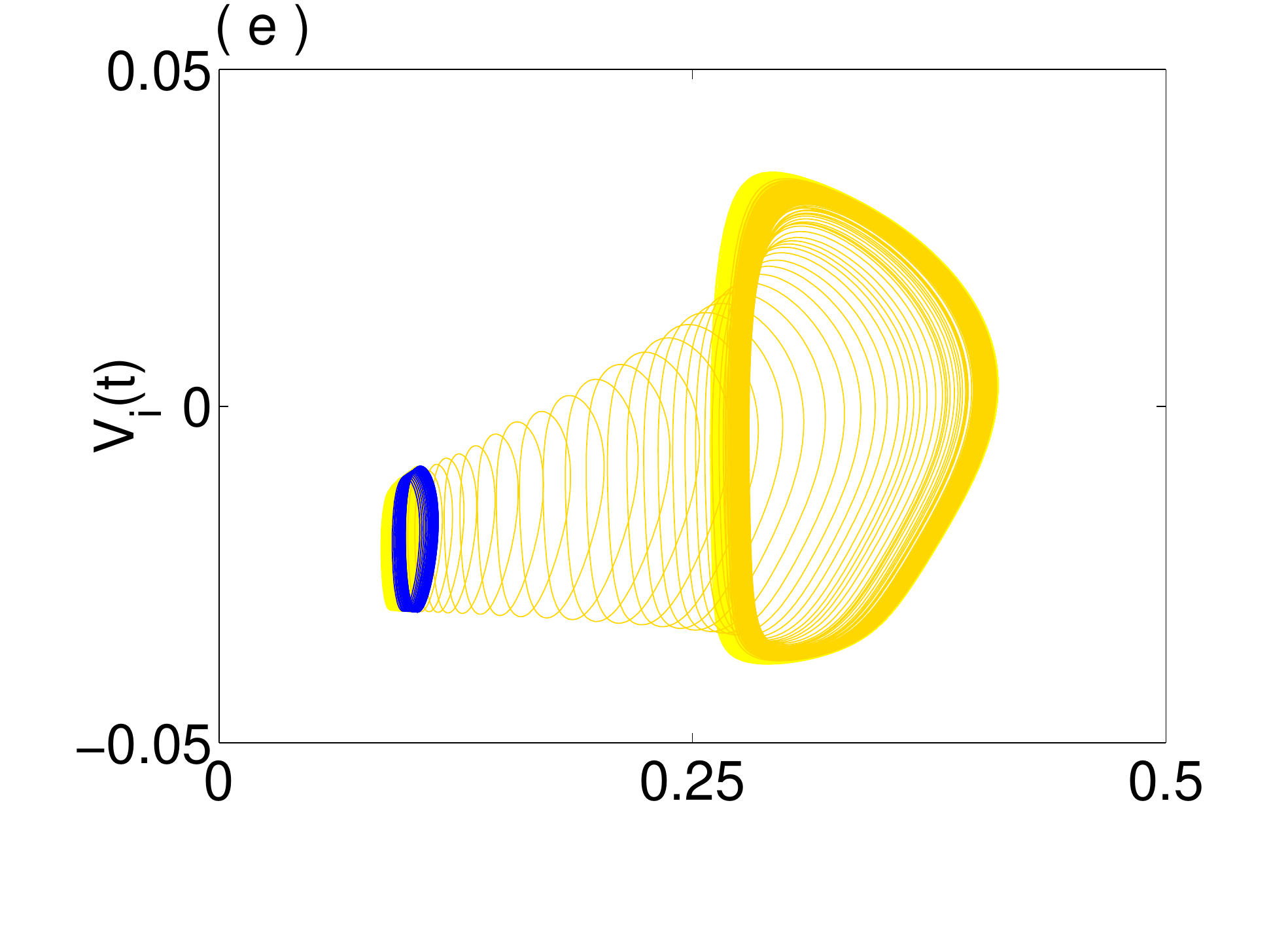}\includegraphics[scale=0.27]{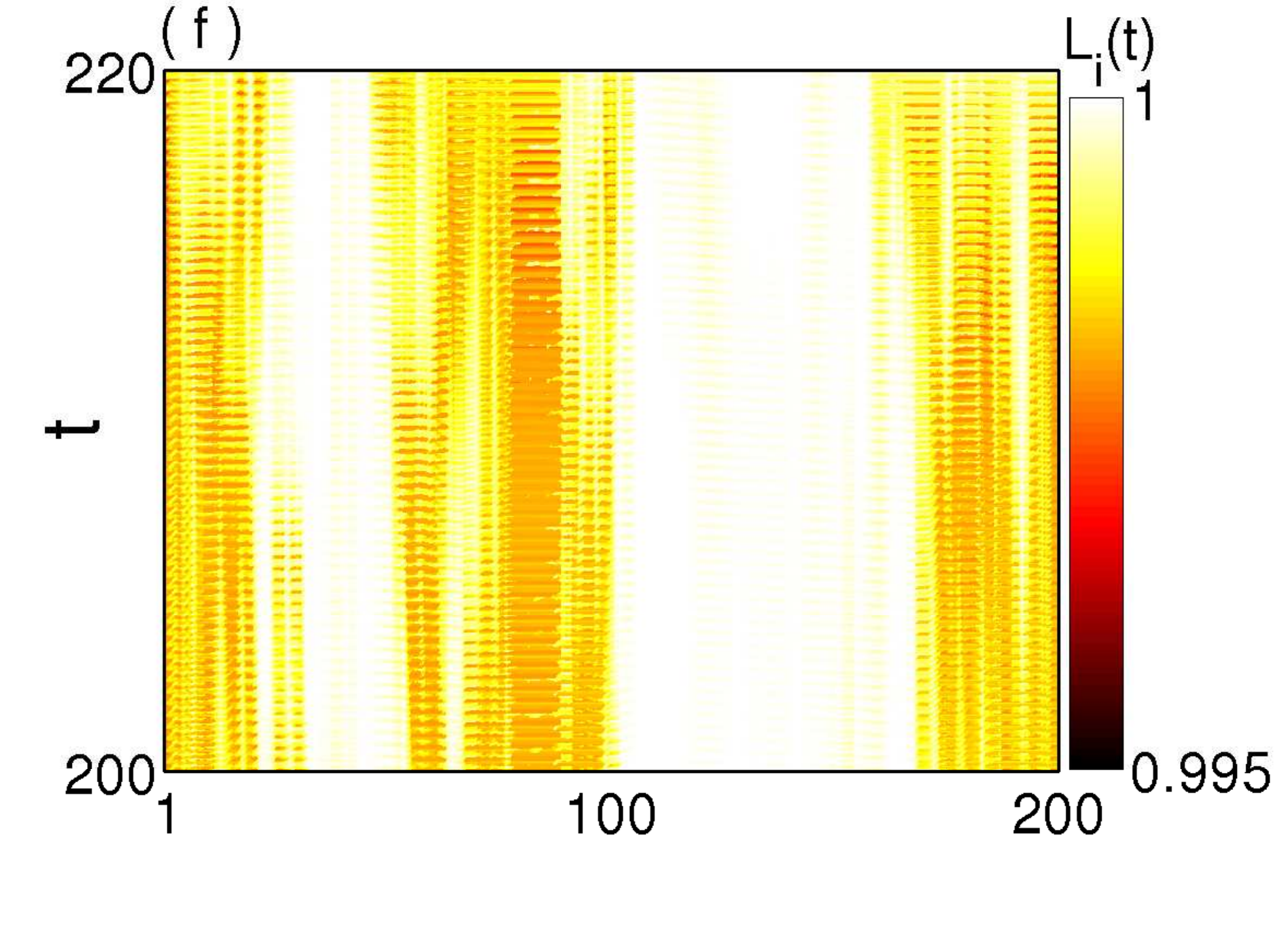}}
\centerline{\includegraphics[scale=0.27]{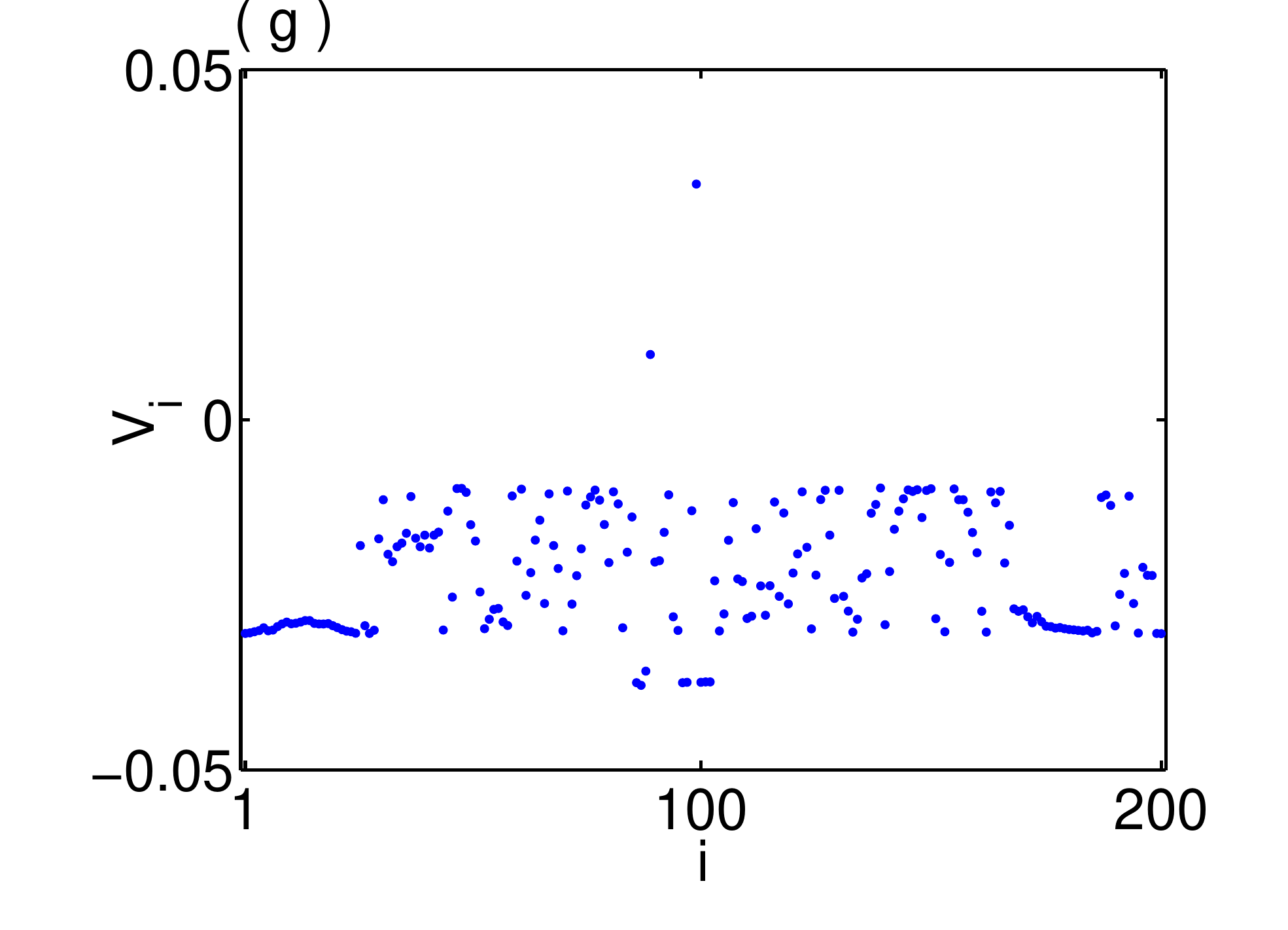}\includegraphics[scale=0.27]{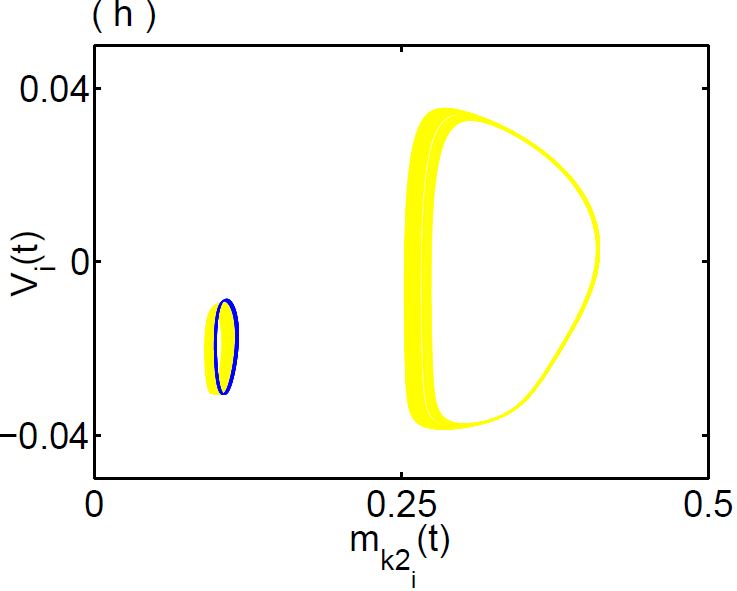}\includegraphics[scale=0.27]{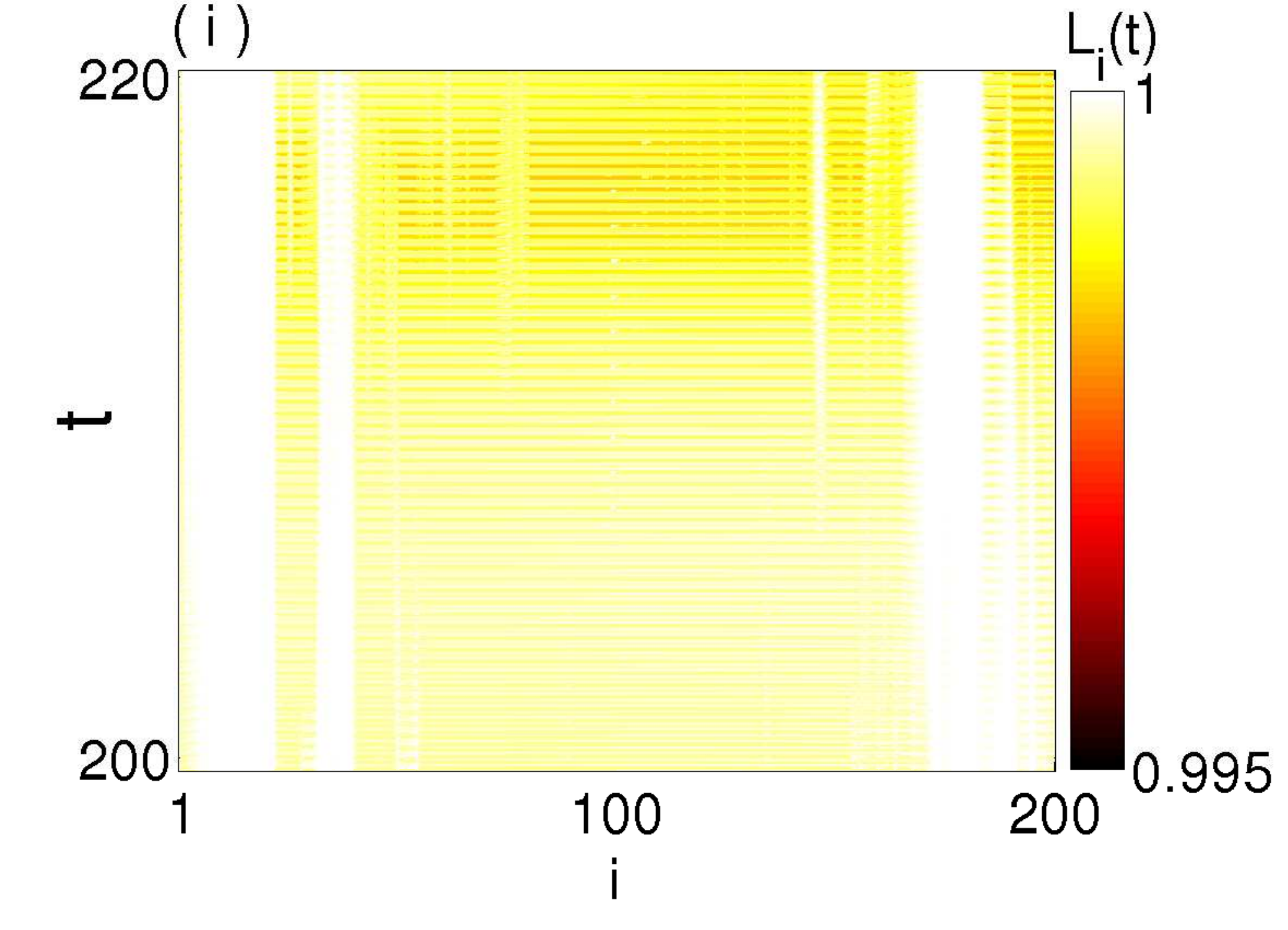}}
\caption{(Color online) Snapshot (first column) of the state variable $V_i$, two-dimensional phase space (second column) in the $(m_{k2_i},V_i)$ plane, and local order parameter $L_i(t)$ (third column). `V' shaped initial conditions are taken from the basin of attraction of the (a,b,c) periodic states PIs with $\epsilon=0.2$; (d,e,f) chaotic states CIs with $\epsilon=0.4$ and (g,h,i) mixed type MIs with $\epsilon=0.3$. Other parameter: $P=20$.}
\label{figure_1}
\end{figure*}

\par Initially for $V^{shift}_{K2}=-0.026~V$ in system (\ref{eq.1}), both tonic spiking modes are represented by two periodic attractors. As the bifurcation parameter $V^{shift}_{K2}$ increases, through a cascade of period-doubling bifurcation, one periodic attractor becomes chaotic. Finally, system (\ref{eq.1}) exhibits the coexistence of a chaotic attractor, and a limit cycle attractor for $V^{shift}_{K2}=-0.025361~V$. These two bistable attractors of a single Leech neuronal oscillator are depicted in Fig. \ref{figure_0}(a). One is a periodic orbit, which is plotted by a red (gray) curve, and the another is a chaotic attractor, shown by a blue (black) curve. Figure \ref{figure_0}(b) demonstrates their basins of attraction, the blue (black) region is the basin of attraction of the chaotic state, while the red (gray) region depicts the basin of attraction of the period-1 limit cycle state. This shows that their basins of attraction are almost well-separated. To get a clearer view of the well-separated basin of attraction, we plot it on the $h_{Na}=0.5$ plane in Fig. \ref{figure_0}(c).

\begin{figure*}
\centerline{\includegraphics[scale=0.55]{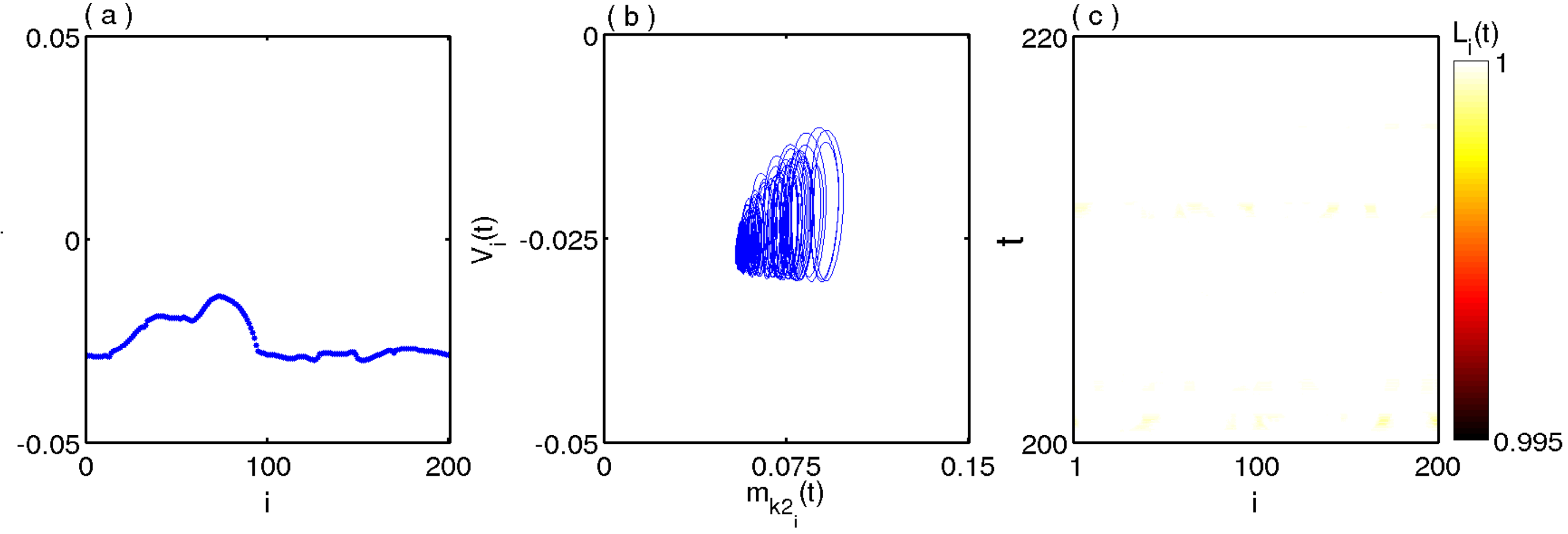}}
\caption{(Color online) (a) Snapshot, (b) 2D phase space and (c) local order parameter of the coherent state for $\epsilon=10.0$ with `V' shaped periodic initial conditions (PIs).}\label{figure_3}
\end{figure*}

\par To construct the neuronal network, a one-dimensional ring of $N$ identical oscillators of the system (\ref{eq.1}) is considered, in which each oscillator is coupled to its $2P$ nearest neighbors. Then the evolution equations of the dynamical network are given as follows:
\begin{equation}
\begin{array}{l}
\label{eq_9}
\dot{V}_{i}=-\frac{1}{C}\big[\bar{g}_{K2}m^2_{K2_i}(V_{i}-E_K )+g_1 (V_{i}-E_1)+\\\bar{g}_{Na}\big(f(A_1,B_1,V_{i})\big)^3 h_{Na_i} (V_{i}-E_{Na} )\big]+\frac{\epsilon}{2P}\sum\limits_{\substack{ j=i-P \\ j\ne i }}^{i+P}(V_{j}-V_{i}),\\
\dot{m}_{K2_i}=\dfrac{f(A_2,B_2+V^{shift}_{K2},V_{i})-m_{K2_i}}{\tau_{K2}} ,\\[7pt]
\dot{h}_{Na_i}=\dfrac{f(A_3,B_3,V_{i})-h_{Na_i}}{\tau_{Na}},
\end {array}
\end{equation}
where $i=1,2,\dots,N$ is the index of the $i^{th}$ oscillator, and the parameter $\epsilon$ is the strength of coupling. For numerical simulation, the number of oscillators $N$ and the number of neighbors in each side $P$ are considered to be fixed at $N=200$ and $P=20$. Throughout this work, we integrate Eq. (\ref{eq_9}) using the fourth-order Runge-Kutta-Fehlberg method with a fixed time step $dt=0.001$ for $3\times10^5$ time iterations. We check that our observations are not specific to either the chosen integrator or the numerical accuracy. In our work, we consider three types of `V' shaped initial conditions (with small random fluctuations) as follows.
\begin{enumerate}
\item PIs: `V' shaped initial conditions for each oscillator are taken from the basin of attraction of the periodic state, i.e., the dotted line in Fig. \ref{figure_0}(c)  as $V_i(0)=0.1-0.2\times\frac{i-1}{\frac{N}{2}-1},~m_{K2_i}(0)=-1.25V_i+0.125,~h_{Na_i}(0)=0.5$ for $1\le i\le\frac{N}{2}$; and $V_i(0)=-0.1+0.2\times\frac{i-\frac{N}{2}-1}{\frac{N}{2}-1},~m_{K2_i}(0)=1.25V_i+0.375,~h_{Na_i}(0)=0.5$ for $\frac{N}{2}+1\le i\le N$.
\item CIs: `V' shaped initial conditions for each oscillator are taken from the basin of attraction of the chaotic state, i.e., the dashed line in Fig. \ref{figure_0}(c) as $V_i(0)=0.1-0.135\times\frac{i-1}{\frac{N}{2}-1},~m_{K2_i}(0)=-0.4444V_i+0.2344,~h_{Na_i}(0)=0.5$ for $1\le i\le\frac{N}{2}$; and $V_i(0)=-0.035+0.135\times\frac{i-\frac{N}{2}-1}{\frac{N}{2}-1},~m_{K2_i}(0)=0.0741V_i+0.2576,~h_{Na_i}(0)=0.5$ for $\frac{N}{2}+1\le i\le N$.
\item MIs: `V' shaped mixed type initial conditions are taken from the entire phase space, i.e., the solid line in Fig. \ref{figure_0}(c) as $V_i(0)=0.1-0.135\times\frac{i-1}{\frac{N}{2}-1},~m_{K2_i}(0)=-0.8519V_i+0.27019,~h_{Na_i}(0)=0.5$ for $1\le i\le\frac{N}{2}$; and $V_i(0)=-0.035+0.135\times\frac{i-\frac{N}{2}-1}{\frac{N}{2}-1},~m_{K2_i}(0)=0.8519V_i+0.3298,~h_{Na_i}(0)=0.5$ for $\frac{N}{2}+1\le i\le N$.
\end{enumerate}
In the next section, our main emphasis will be to investigate the emergence of chimera states and the effects of these three types of initial conditions on them.

\section{Emergence of chimera states in a Leech neuronal network}\label{chimera}
We first investigate the emergence of chimera states in the non-locally coupled Leech neuronal network, Eq. (\ref{eq_9}), by varying the interaction strength $\epsilon$. The first, second, and third columns of Fig. \ref{figure_1} respectively show the snapshot, two-dimensional phase space in the $(m_{k2_i},V_i)$ plane, and local order parameter $L_i$ (which is defined later). In the top panel of Fig. \ref{figure_1}, initial conditions are taken from PIs, with coupling strength $\epsilon=0.2$. The scatter population in the right side of Fig. \ref{figure_1}(a) indicates that the corresponding neurons are drifting relative to the amplitude locked oscillators. Here, the neurons from $i=1$ to $i=116$ spike coherently, while the remaining neurons spike with different frequencies. From the intrinsic periodic state, all the neurons exhibit the quasi-periodic state depicted in Fig. \ref{figure_1}(b), which is lying near that periodic state. Here, the oscillators corresponding to the yellow curves are in the incoherent state, and all the oscillators from the coherent region follow the blue trajectories. The yellow trajectories of the incoherent oscillators are more spread out in the phase space compared to the those of coherent oscillators. The middle panel of Fig. \ref{figure_1} is drawn for CIs initial conditions, where $\epsilon$ is fixed at $0.4$. Figure \ref{figure_1}(d) represents the snapshot, where the oscillators from $i=103$ to $i=171$ maintain the coherent state. Corresponding two-dimensional phase space is plotted in Fig. \ref{figure_1}(e), where the blue trajectories are for the coherent group of oscillators, while the yellow curves represent the trajectories of the incoherent group of oscillators. Here, CIs-type initial conditions are chosen. But due to the interaction, the trajectories of a few neurons spike quasi-periodically, while the rest of the neurons remain in the chaotic state, and one neuron is oscillating from the chaotic to the quasi-periodic state (deep yellow). For MIs initial conditions, the snapshot of the chimera states is drawn in Fig. \ref{figure_1}(g) for $\epsilon=0.3$. In one domain, the oscillators are desynchronized, and the another two domains, they are in coherent motion. As in the previous case, all the coherent oscillators follow a quasi-periodic trajectory (blue), but the incoherent oscillators are divided into two groups, where one of them is in the chaotic state while the other follows the quasi-periodic state.

\par By using the notion of a local order parameter, we characterize the spatial coexistence of coherent-incoherent patterns of the chimera states. To determine the coherent-incoherent clusters and chimera states, the concept of a local order parameter is often used \cite{r46}. The local order parameter demonstrates the neighbor ordering of the neuronal network and quantifies the coherent and incoherent clusters. This parameter is defined as
\begin{equation}
 \begin{array}{l}\label{eq.7}
  L_{i}(t)=\bigg| \dfrac{1}{2\delta}\sum\limits_{|i-k| \le \delta}e^{\phi_{k}(t)\sqrt{-1}} \bigg|,~~~~~~~i=1,2,\cdots,N,
 \end{array}
\end{equation}
where $\delta$ denotes the window size of the $i^{th}$ neuron used for the spatial average; here we consider $\delta=3$. The geometric phase of the $k^{th}$ neuron is determined by the formula
\begin{equation}
 \begin{array}{l}\label{eq.8}
  \phi_k(t)=\arctan\Big(\frac{y_k(t)}{x_k(t)}\Big).
 \end{array}
\end{equation}
The color bars in the third column of Fig. (\ref{figure_1}) represent the variation of $L_i(t)$ for three different types of initial conditions, which corroborate with the corresponding snapshot plots in the first column. White region corresponds to the values of $L_i(t)$ equal to $1$, which indicate the maximum coherency of the $i^{th}$ neuron, and it belongs to the coherent population. However, for the $i^{th}$ neuron that belongs to the incoherent domain of the chimera state, $L_i$ is obtained at less than $1$, which is shown as other colors in the color bars.

\par When the coupling strength gets higher values than $\epsilon=8.0$, the chimera states disappear, and the network of neurons tends to oscillate coherently. Figure \ref{figure_3}(a) shows the snapshot of the coherent state, taking the `V' shaped periodic initial conditions for the coupling strength $\epsilon=10.0$. This depicts a smooth coherent manifold among the oscillators. In this coherent state, all the oscillators follow a chaotic motion near its intrinsic periodic orbit, drawn in Fig. \ref{figure_3}(b). At the chimera states, the neuronal network (\ref{eq_9}) possesses different types of spiking resting behavior as we observe from the phase space plots of Fig. (\ref{figure_1}) (second column). But at the coherent state, the neuronal network exhibits the same chaotic trajectory regardless of the other two types of initial conditions (checked separately, but the results are not presented here). The corresponding local order parameter is plotted in Fig. \ref{figure_3}(c). Throughout the time interval $t\in[200,220]$, $L_i(t)\simeq 1$ for all $i=1,2,\dots,N$. This describes the local coherency of all the oscillators. From these results, it is clear that the oscillators in the coherent states are amplitude-locked, and all of them oscillate at an identical instantaneous frequency and smooth amplitude profile.

\section{Analytical study of the chimera to coherent transition}\label{analysis}
\par In chimera states, symmetrically coupled identical oscillators are split into two regions; one is a coherent region where the phases of the oscillators are locked, and the other is an incoherent region where the oscillators are in a drifting motion. So, in the chimera states, there exists a subpopulation of oscillators that oscillate in a drifting manner, which is random. Any oscillator from this incoherent domain, could not obey a smooth transformation of the remaining oscillators. Therefore, in the whole network, there could not exist a smooth transformation function of a particular oscillator in terms of all other oscillators. But for coherent states, the amplitudes and phases are locked for all the oscillators. At every time, their snapshots show a smooth profile. From this fact, we claim that for all $i=1,2,\dots,N$, there exists a smooth transformation $\Phi:\mathbb{R}^{d(N-1)}\rightarrow\mathbb{R}^d$ such that
\begin{equation}
\begin{array}{l}\label{eq_8}
{\bf x}_i=\Phi({\bf x}_1,{\bf x}_2,\dots,{\bf x}_{i-1},{\bf x}_{i+1},\dots,{\bf x}_N).
\end {array}
\end{equation}
In other words, we assert that in the coherent state, the network of oscillators follows generalized synchronization, which is a general type of correlation among the coupled oscillators. 

\par If there exists such a smooth mapping $\Phi$, and a manifold $\mathcal{M}=\big\{({\bf x}_1,{\bf x}_2,\dots,{\bf x}_N)~\big|~{\bf x}_i=\Phi({\bf x}_1,{\bf x}_2,\dots,{\bf x}_{i-1},{\bf x}_{i+1},\dots,{\bf x}_N)\big\} \subset \mathbb{R}^{dN}$, such that the trajectories of network (\ref{eq_1}) with random initial conditions from its basin of attractions approach $\mathcal{M}$ as $t\rightarrow\infty$, then the network of oscillators is said to realize generalized synchronization. Here $\mathcal{M}$ is the generalized synchronization manifold, which is an invariant manifold, and it is attracting if the generalized synchronization state is stable. The existence of the smooth mapping $\Phi$ guarantees the functional relationship of ${\bf x}_i$ with ${\bf x}_1,{\bf x}_2,\dots,{\bf x}_{i-1},{\bf x}_{i+1},\dots,{\bf x}_N$.

\par To mathematically detect the occurrence of the generalized synchronization states, we use the auxiliary-system approach \cite{ausys_1}. It is a powerful mathematical tool to detect the above-mentioned functional relationship among coupled oscillators. This approach is applied to bidirectionally coupled dynamical networks by introducing an auxiliary node system corresponding to each original node \cite{ausys_2}. It can be described for the $i^{th}$ node as
\begin{equation}
\begin{array}{l}
\label{eq_4}
\begin{array}{lll}
\dot{\bf y}_i=F({\bf y}_i)+\dfrac{\epsilon}{d(i)}\sum\limits_{j=1}^{N}\mathscr{A}_{ij}\Gamma({\bf x}_j-{\bf y}_i).
\end{array}
\end{array}
\end{equation}
If ${\bf y}_i$ identically synchronizes with ${\bf x}_i$, then the GS of the network (\ref{eq_1}) is assumed to occur. Obviously, all the nodes show identical behavior with their auxiliary counterpart. The system (\ref{eq_1}) is said to achieve generalized synchronization if $\norm{{\bf x}_i(t)-{\bf y}_i(t)}\rightarrow 0$ as $t\rightarrow\infty$, where $\norm{\cdot}$ is the Euclidean norm. Considering the variation of the $i^{th}$ node ${\bf z}_i(t)={\bf y}_i(t)-{\bf x}_i(t)$, we can get the equation of motion of the error systems as
\begin{equation}
\begin{array}{l}
\label{eq_10}
\dot{\bf z}_i=F({\bf x}_i+{\bf z}_i)-F({\bf x}_i)-\epsilon\Gamma {\bf z}_i,\hspace{10pt}i=1,2,\dots,N.
\end{array}
\end{equation}

\begin{figure}
\centerline{\includegraphics[scale=0.5]{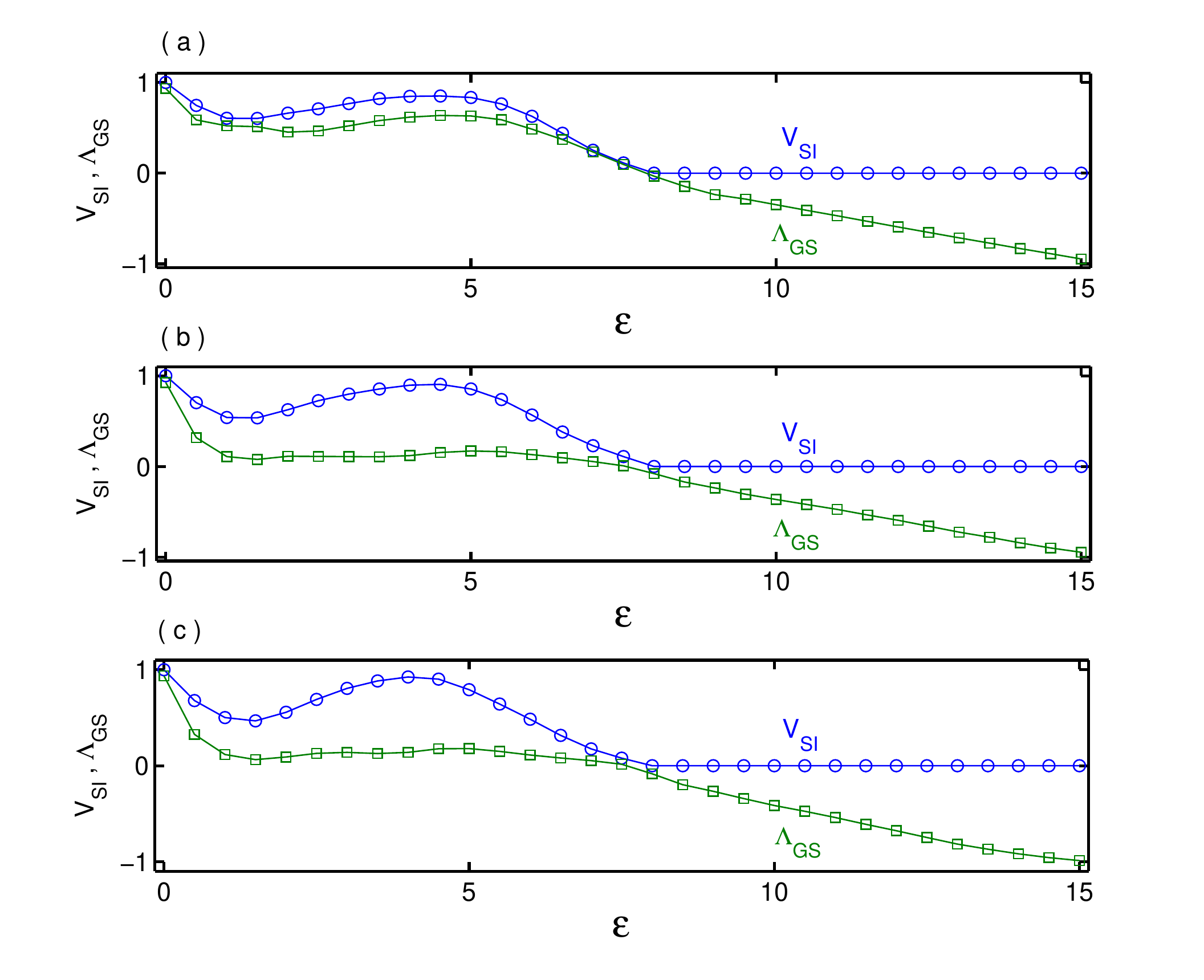}}
\caption{(Color online) Variation of $V_{\mbox{SI}}$ (blue circle curve) and $\Lambda_{\mbox{GS}}$ (red square curve) with respect to $\epsilon$ for (a) periodic initial conditions, (b) chaotic initial conditions and (c) mixed initial conditions.}
\label{figure_4}
\end{figure}

\par For the local stability of the generalized synchronization manifold, we consider a small difference ${\bf z}_i(t)$ of the state variables of the original system and its auxiliary partner. Then we get the coherent stability equation as
\begin{equation}
\begin{array}{l}
\label{eq_5}
\dot{\bf z}_i=[JF({\bf x}_i)-\epsilon\Gamma] {\bf z}_i,\hspace{20pt}i=1,2,\dots,N.
\end{array}
\end{equation}
Here $J$ denotes the Jacobian of the function with respect to the arguments, and $({\bf x}_1,{\bf x}_2,\dots,{\bf x}_N)$ is the state variable of the generalized synchronization manifold obeying Eq. (\ref{eq_1}). Now,
\begin{equation}
\begin{array}{l}
\norm{{\bf z}_i}\rightarrow 0 \Leftrightarrow~\norm{{\bf x}_i-{\bf y}_i}\rightarrow 0 \mbox{ as } t\rightarrow\infty.
\end{array}
\end{equation}
Therefore, generalized synchronization of the network (\ref{eq_1}) is achieved if and only if the trivial solution of the above error system (\ref{eq_5}) is asymptotically stable. The linearized Eq. (\ref{eq_5}) will be solved in parallel to the original non-linear Eq. (\ref{eq_1}) for ${\bf x}_i$, which then allows us to compute all its Lyapunov exponents. The coherent stability function $\Lambda_{\mbox{GS}}$, the maximum of these Lyapunov exponents as a function of $P$ and $\epsilon$, actually gives the necessary condition for the stability of the chimera and coherent states. On the basis of the  linear stability theory, the negative Lyapunov exponents can bring about the stability of the trivial solution of the error system. So, the maximum Lyapunov exponent, $\Lambda_{\mbox{GS}}$, of the system (\ref{eq_5}) should be negative for the local stability of the coherent states. Since the chimera and the coherent states are mutually exclusive and exhaustive, $\Lambda_{\mbox{GS}}>0$ implies the stability of the chimera states, while $\Lambda_{\mbox{GS}}<0$ indicates the stability of the coherent states. The coherent stability function allows one to quickly establish whether the coherent state is stable or not. In this way, it reveals the instability of the chimera states. Here, $\Lambda_{\mbox{GS}}>0$ is the necessary condition for the chimera states, but it is not sufficient, as it may also exhibit an incoherent state.

\par Finally, for the non-locally coupled Leech neuronal network, our required coherent stability equation can be written as
\begin{widetext}
\begin{equation}
\begin{array}{l}
\label{eq_11}
~~~\dot{z}_i^{V}=-\frac{1}{C}\Big[	 \bar{g}_{K2}m^2_{K2_i}+g_1+\bar{g}_{Na}\big(f(A_1,B_1,V_{i})\big)^3 h_{Na_i}+3\bar{g}_{Na}\big(f(A_1,B_1,V_{i})\big)^2f_V(A_1,B_1,V_{i})(V_{i}-E_1)h_{Na_i}	 \Big]z_i^{V}\\[5pt]\hspace{36pt} -\frac{1}{C}\Big[2\bar{g}_{K2}m_{K2_i}(V_{i}-E_K )z_i^{m_{K2}}+\bar{g}_{Na}\big(f(A_1,B_1,V_{i})\big)^3(V_{i}-E_{Na})z_i^{h_{Na}}	\Big]-\epsilon z_i^{V},\\[8pt]
\dot{z}_i^{m_{K2}}=\dfrac{f_V(A_2,B_2+V^{shift}_{K2},V_{i})z_i^{V}-z_i^{m_{K2}}}{\tau_{K2}},\\[8pt]
~\dot{z}_i^{h_{Na}}=\dfrac{f_V(A_3,B_3,V_{i})z_i^{V}-z_i^{h_{Na}}}{\tau_{Na}},\hspace{50pt} i=1,2,\dots,N.
\end{array}
\end{equation}
\end{widetext}
Here, $f_V(A,B,V)=\frac{\partial}{\partial V}f(A,B,V)=\frac{-A\exp[A(B+V)]}{\big\{1+\exp[A(B+V)]\big\}^2}$, and the state variables $(V_i,m_{K2_i},h_{Na_i})$ of the generalized synchronization are manifold dominated by Eq. (\ref{eq_9}). When the chimera states of the non-locally coupled system become unstable, the coherent states turn out to be stable. The maximum Lyapunov exponent, $\Lambda_{\mbox{GS}}$, of the above error system (\ref{eq_11}) becomes negative, and with the stable coherent state, all the oscillators evolve generalized synchronously.

\begin{figure}
\centerline{\includegraphics[scale=0.34]{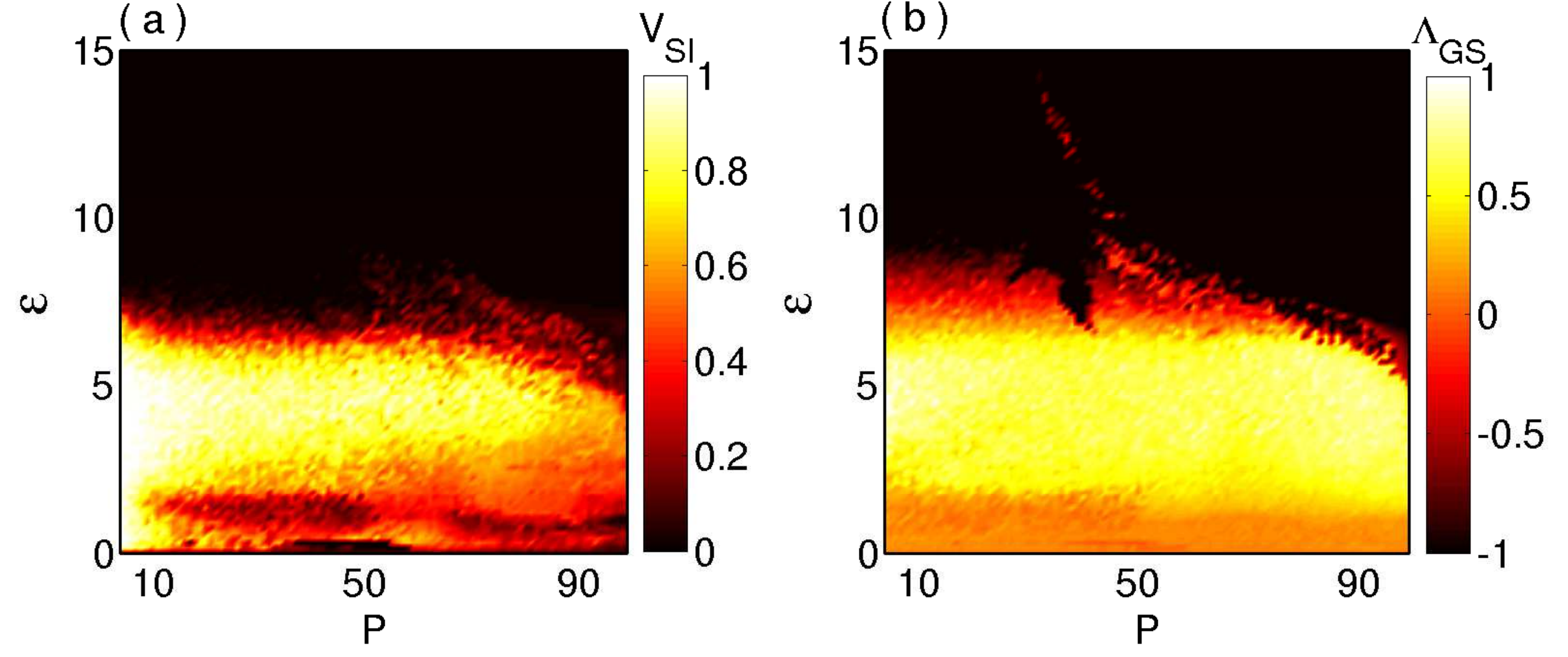}}
\caption{(Color online) Two-parameter phase diagram in $(P,\epsilon)$ plane with periodic initial conditions. Color bars represent the variation of (a) $V_{\mbox{SI}}$ and (b) $\Lambda_{\mbox{GS}}$.}
\label{figure_5}
\end{figure}

\par To distinguish the chimera states from the other states, such as incoherent and coherent states, the time-averaged statistical measure of the strength of incoherence $(V_{\mbox{SI}})$ is widely used \cite{r34}. To compute the value of $V_{\mbox{SI}}$, we divide the total number of oscillators into an even number of $M$ bins, each of which is of equal length $n=\frac{N}{M}$. Then the local standard deviation $\sigma(m)$ is defined as
\begin{equation}
\begin{array}{l}\label{eq_12}
\sigma(m)=\left\langle\sqrt{\frac{1}{n}\sum\limits_{j=n(m-1)+1}^{nm}\big[z_j-\langle z\rangle\big]^2}\right\rangle_t,
\end{array}
\end{equation}
where $m=1,2,\dots,M$; the difference of the potential state variables $V_i$ is $z_i$, which is defined as $z_i=V_i-V_{i+1}$ for $i=1,2,\dots,N$ and $V_{N+1}=V_1$. This new state variable better illustrates the occurrence of the different correlation state in the coupled systems. When the two neighbours, namely the $i^{th}$ and $(i+1)^{th}$ oscillators, oscillate coherently, the value of $z_i$ will be near zero. If the two neighboring oscillators, $i^{th}$ and $(i+1)^{th}$, evolve incoherently, $z_i$ will take a value much greater than zero. Here $\langle\cdot\rangle_t$ denotes the long time average, and $\langle z\rangle=\frac{1}{N}\sum\limits_{i=1}^{N}z_i$. Then the strength of incoherence is defined as
\begin{equation}
\begin{array}{l}\label{eq_13}
V_{\mbox{SI}}=1-\frac{1}{M}\sum\limits_{m=1}^{M}\Theta(\delta-\sigma(m)),
\end{array}
\end{equation}
where $\delta$ is the pre-defined threshold. In our work we choose $\delta=0.0025$, and $\Theta(\cdot)$ is the Heaviside step function. Here, $V_{\mbox{SI}}=1$ and $V_{\mbox{SI}}=0$ respectively represent the incoherent and coherent states, while the chimera states are identified by the values of $V_{\mbox{SI}}$ belonging in the interval $(0,1)$.

\par The variation of the maximum Lyapunov exponent $\Lambda_{\mbox{GS}}$ of the error systems (\ref{eq_11}) is shown in Fig. (\ref{figure_4}) with respect to $\epsilon$. For the coupling strength $\epsilon\ge8.0$, the zero solution of the system (\ref{eq_11}) becomes stable, and accordingly, each node of the original system (\ref{eq_9}) is synchronized with its counterpart node in the auxiliary system (\ref{eq_4}). Here the coherent state is achieved together with the generalized synchronization state. To verify our analytical conditions, we plot $V_{\mbox{SI}}$ together with $\Lambda_{\mbox{GS}}$ by varying $\epsilon$. The top, middle, and bottom panels are, respectively, for PIs-, CIs-, and MIs-types initial conditions. The values of $V_{\mbox{SI}}$ are plotted by blue circle curves, while $\Lambda_{\mbox{GS}}$ are plotted by green square curves. For all three types of initial conditions, as the values of $V_{\mbox{SI}}$ abruptly drops from $1$, it is indicated for a slight increment of $\epsilon$ from zero that the incoherent states become unstable and chimera states appear. The values of $V_{\mbox{SI}}$ remain nonzero up to $\epsilon<8.0$; in that range of $\epsilon$ the values of $\Lambda_{\mbox{GS}}$ remains non-negative. From $\epsilon=8.0$, as soon as $V_{\mbox{SI}}$ becomes zero, $\Lambda_{\mbox{GS}}$ turns out to be negative. That is, for $\epsilon\ge8$, the chimera states become unstable and coherent states become stable. For still higher values of $\epsilon$, $V_{\mbox{SI}}$ remains zero and $\Lambda_{\mbox{GS}}$ remains negative, which indicates the persistence of the coherent states. In conclusion, for all three types of initial conditions, our analytical study exactly matches the numerical simulations. This analysis provides a straightforward and precise characterization of the chimera to the coherent transition in coupled dynamical systems, probably irrespective of the complexity of the underlying network architecture and system dynamics.

\par To explore the complete scenario of the variation of $V_{\mbox{SI}}$ and $\Lambda_{\mbox{GS}}$, we compute the phase diagram by simultaneously varying $P$ and $\epsilon$ in the range of $P\in[1,99]\cap\mathbb{N}$ and $\epsilon\in[0,15]$. In the $(P,\epsilon)$ parameter space of a non-locally coupled Leech neuronal oscillator, the variations of $V_{\mbox{SI}}$ and $\Lambda_{\mbox{GS}}$ are drawn in Figs. \ref{figure_5}(a) and \ref{figure_5}(b) respectively. For this result, the PIs' initial conditions are chosen. For the other two types of initial conditions (CIs and MIs), almost similar results are presented in the Appendix. From Fig. \ref{figure_5}(a), it is noted that for lower values of $P$, the incoherent states persist up to $\epsilon\simeq5.5$. Beyond that, chimera states emerge. But for higher values of $P$, chimera states appear as soon as we switch on the interaction. For the incoherent and chimera regions, the values of $\Lambda_{\mbox{GS}}$ remain non-negative in Fig. \ref{figure_5}(b). For larger values of $\epsilon$, all nodes oscillate coherently, irrespective of the values of $P$. It emphasizes that the coupling strength plays an important role in changing the dynamical properties of the network. For higher values of $\epsilon$, as soon as chimera states are annihilated and coherent states emerge, the values of $\Lambda_{\mbox{GS}}$ become negative together with $V_{\mbox{SI}}=0$.

\begin{figure}[b]
\centerline{\includegraphics[scale=0.35]{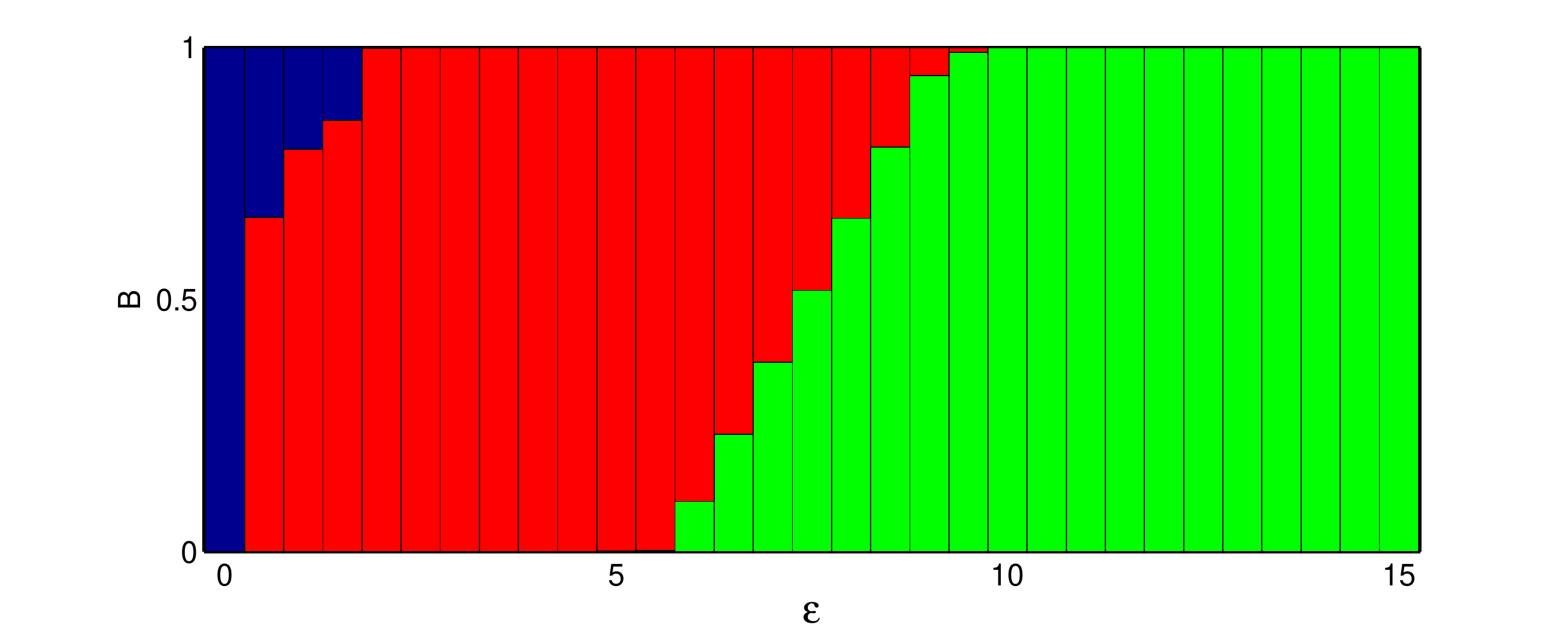}}
\caption{(Color online) The variation of basin stability of incoherent [blue (black) bars], chimera [red (dark gray) bars] and coherent states [green (gray) bars] with respect to $\epsilon$ for $P=20$.}
\label{figure_6}
\end{figure}

\section{Global stability of the chimera and coherent states}\label{global}

\par From the discovery of the chimera states, the initial conditions play a crucial role. For many coupled systems, chimera states emerge only for a few specific initial conditions \cite{cmra_bs1,cmra_srep}. Especially in multi-stable systems, the emergence of chimera states depends on the initial conditions. Since the isolated Leech neuronal model itself possesses multi-stable behavior, we now investigate the robustness of the initial conditions on the chimera states using basin stability measurement \cite{cmra_srep,bs_delay}. This is a non-linear, non-local approach, that can be easily applied to high-dimensional multi-stable systems. It is closely related to the volume of the basin of attraction. The basin stability tells us how stable the incoherent, chimera, and coherent states are against multifarious initial conditions. Recently, this measurement was used to quantify the multi-stable incoherent, chimera, and coherent states in non-locally coupled time-delayed Mackey-Glass oscillators \cite{cmra_srep}. Here we have characterized different dynamical states by calculating the strength of incoherence. Now our main emphasis will be to identify the variation of basin stability in the parameter range of coupling strength $\epsilon$.

\par For numerical simulation, we choose random initial conditions from the phase-space volume $[-0.05,0.05]\times[0.0,0.5]\times[0.0,1.0]$ of the Leech oscillator. Then we integrate the entire system (\ref{eq_9}) for $Q$ (sufficiently large) different initial conditions, for which $T$ initial conditions finally arrive at our desired state. Then the basin stability $(B)$ of that particular state can be estimated as $\frac{T}{Q}$. The value $B\in[0,1]$, where $B=0$ indicates the instability of that state, and it is globally stable only if $B=1$. $B\in(0,1)$ indicates the coexistence of more than one dynamical state. Actually, basin stability corresponds to the probability of getting the state for any random initial conditions in the classical sense.

\par Now we investigate the probability of the emergence of the chimera state together with the incoherent and coherent states. The variations of $B$ for incoherent, chimera, and coherent states with respect to the coupling strength $\epsilon$ are plotted in Fig. (\ref{figure_6}). Here blue (black), red (dark gray), and green (gray) bars respectively represent the incoherent, chimera, and coherent states. For the values of $\epsilon$ almost near zero, the incoherent state solely  dominates. By increasing the values of $\epsilon$, the $B$ of the chimera states abruptly increases, while the incoherent states are annihilated as well. From $\epsilon=2.0$, the chimera states become mono-stable, and they remain so up to $\epsilon=5.0$. Beyond that value of $\epsilon$, the basin stability of the chimera states gradually decreases and that of the coherent states improves. At $\epsilon=10$, the basin stability of the coherent states reaches $1$ and acquires full space in the basin volume, which is maintained for still higher values of $\epsilon$. Such changes of $B$ give an indication of the annihilation of incoherent and chimera states and the global stability of coherent states by varying the coupling strength.

\begin{figure*} \centerline{\includegraphics[scale=0.4]{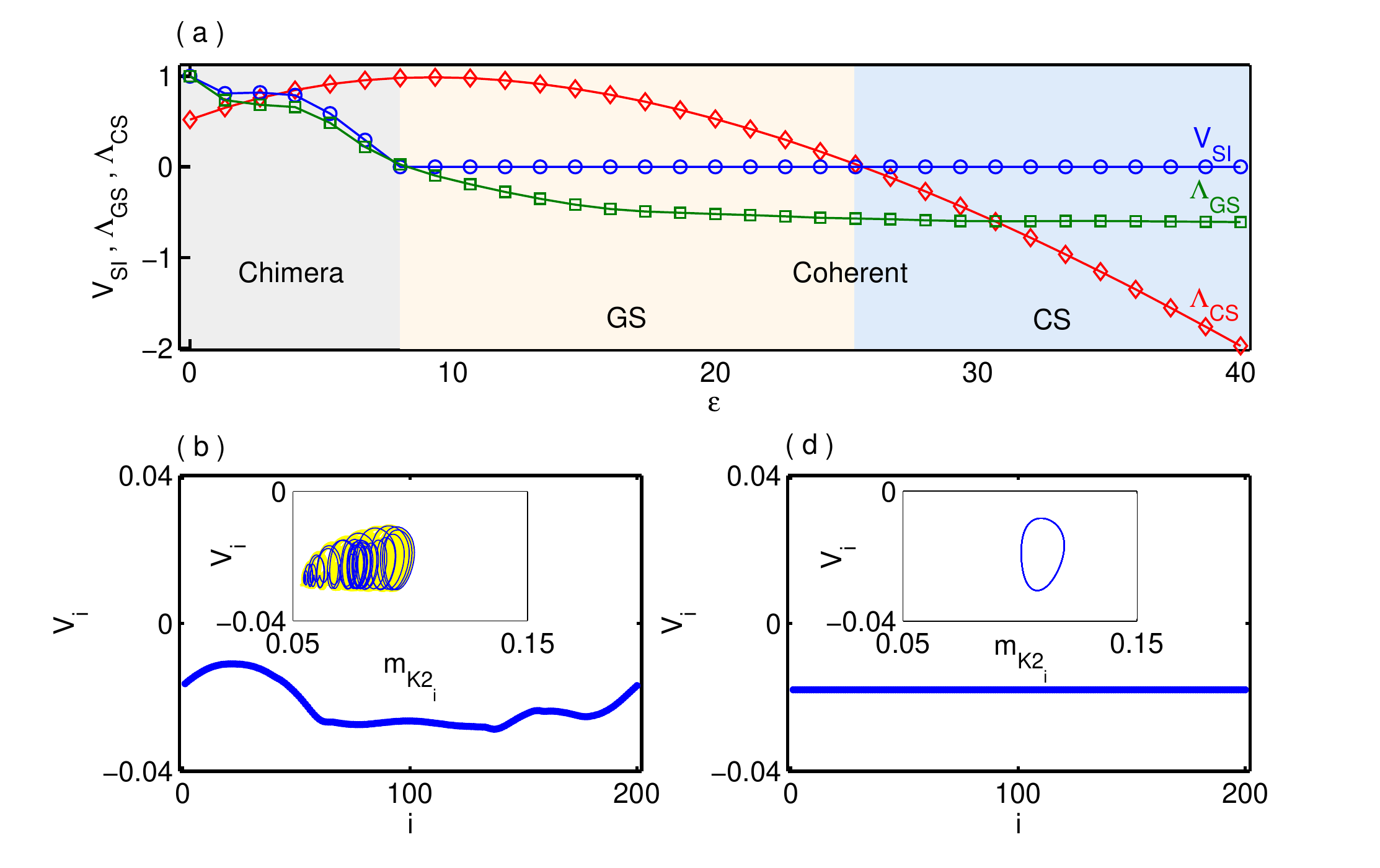}\includegraphics[scale=0.4]{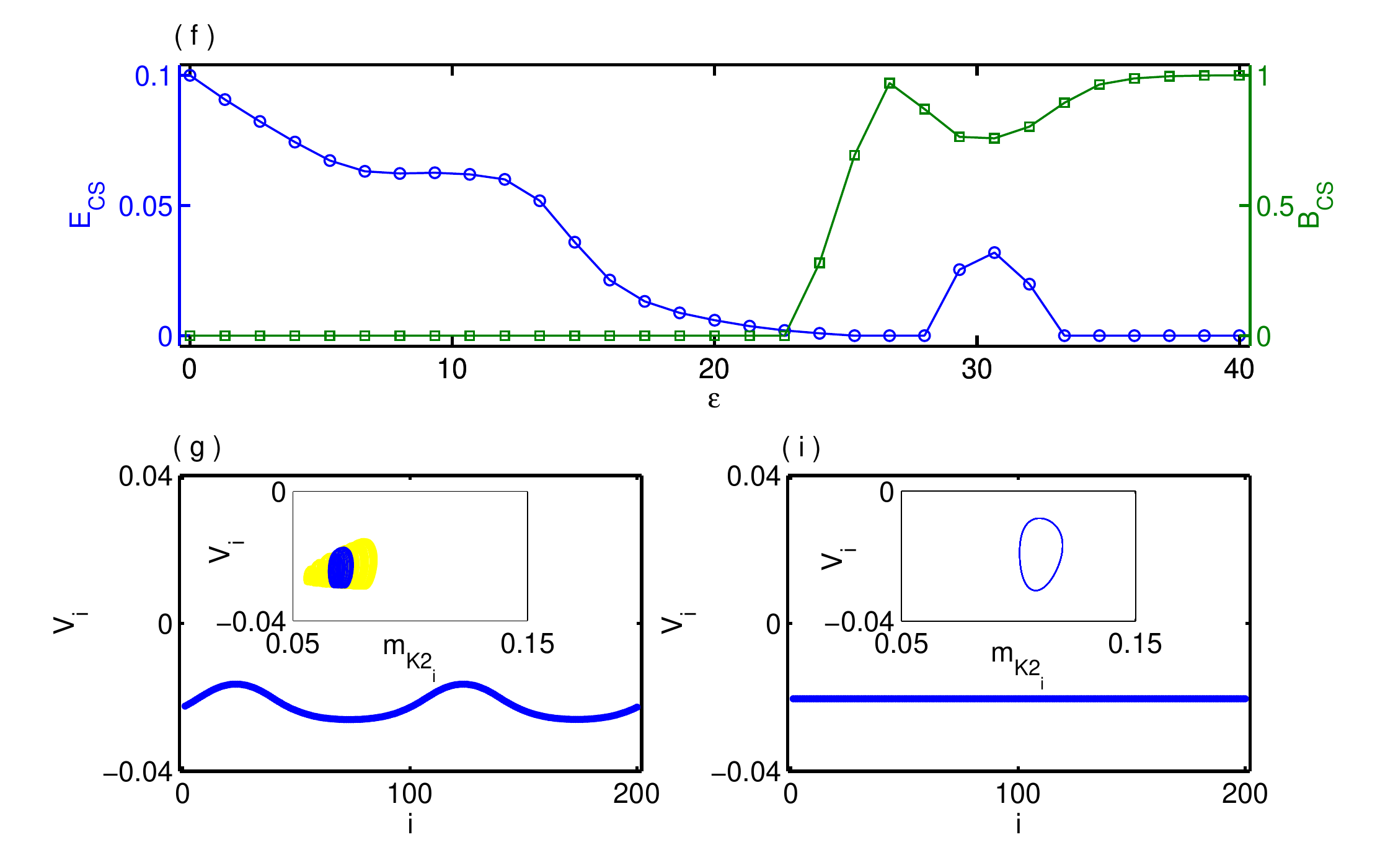}}
\centerline{\hspace{8pt}\includegraphics[scale=0.225]{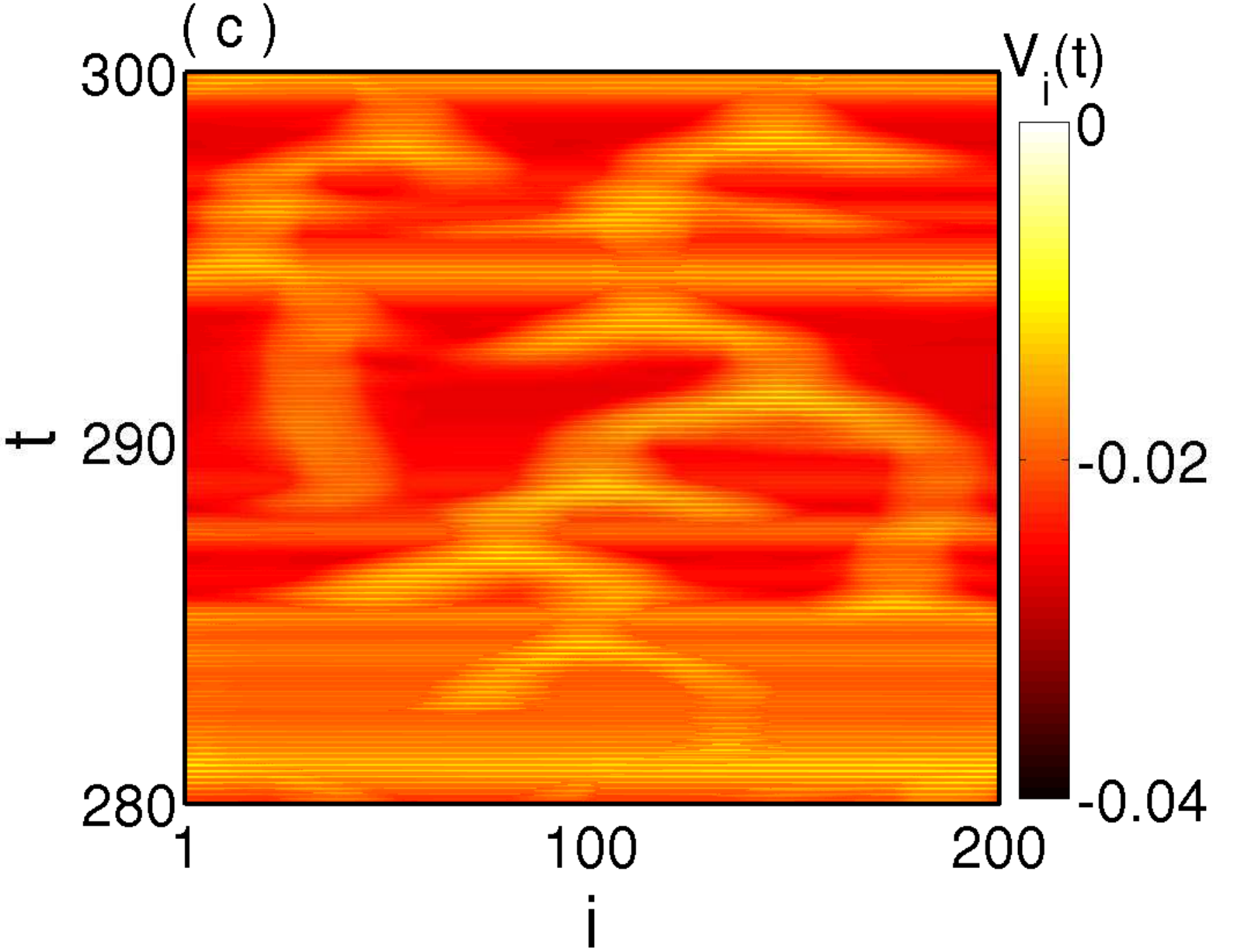}\includegraphics[scale=0.225]{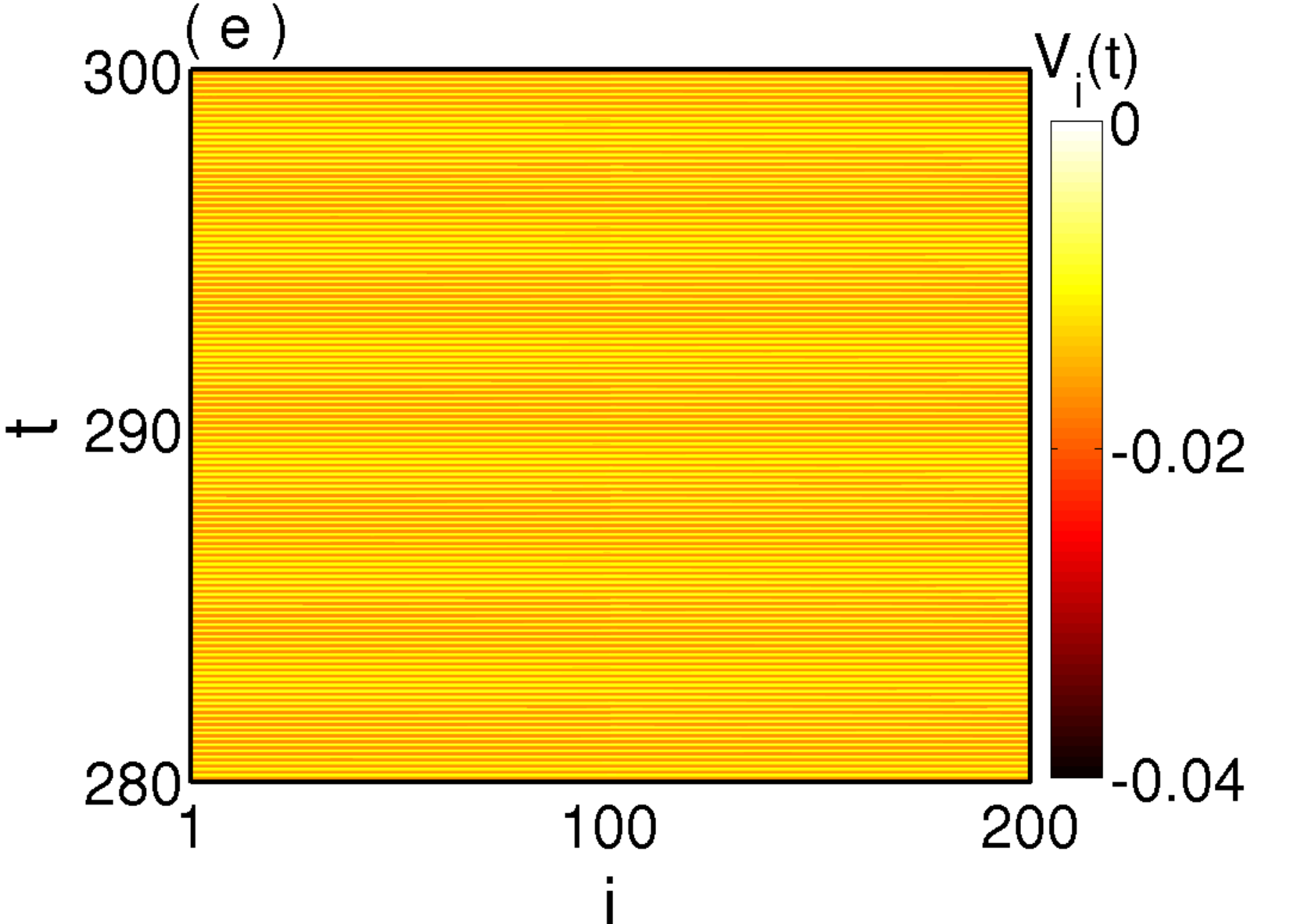}\hspace{27pt}\includegraphics[scale=0.225]{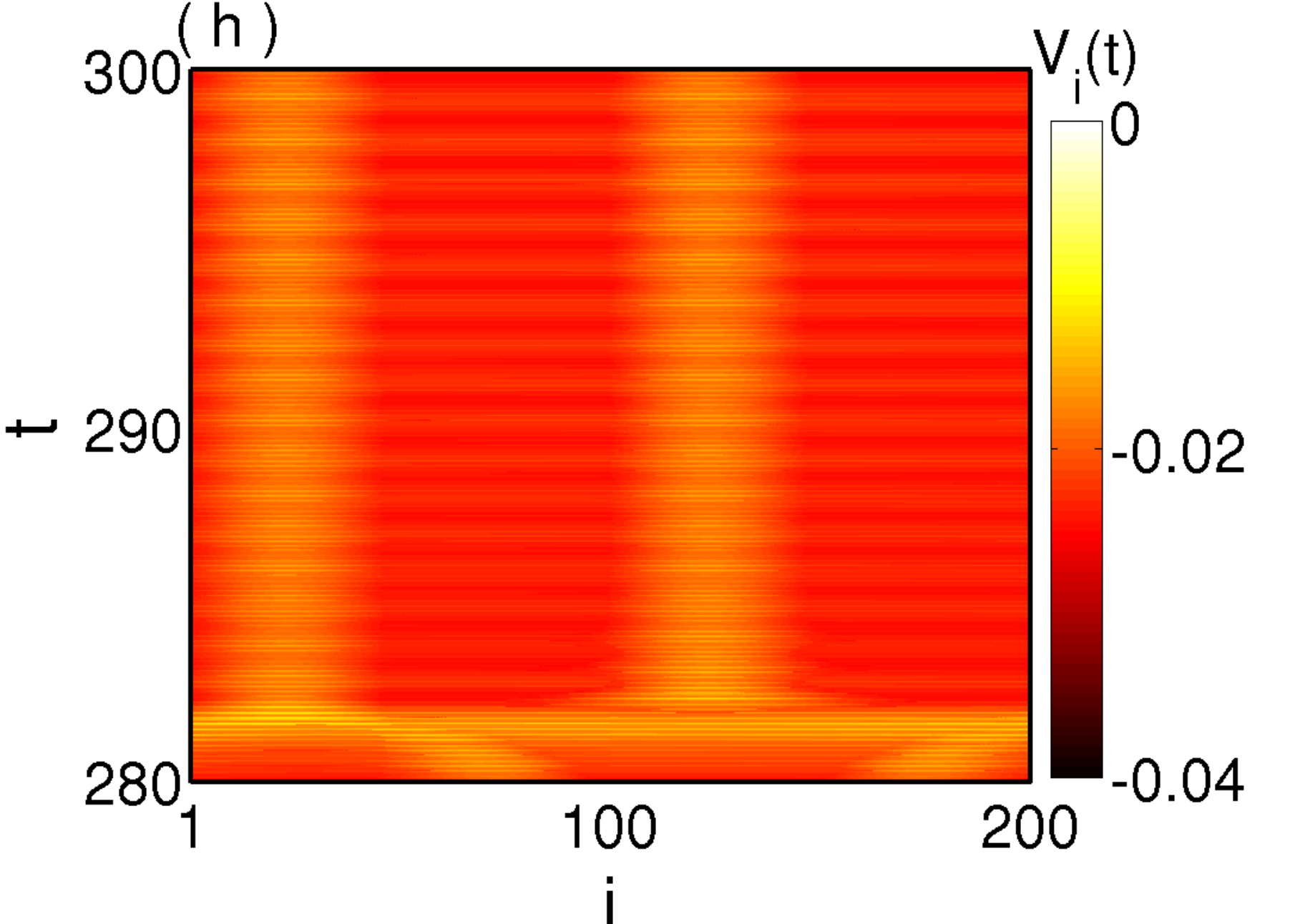}\includegraphics[scale=0.225]{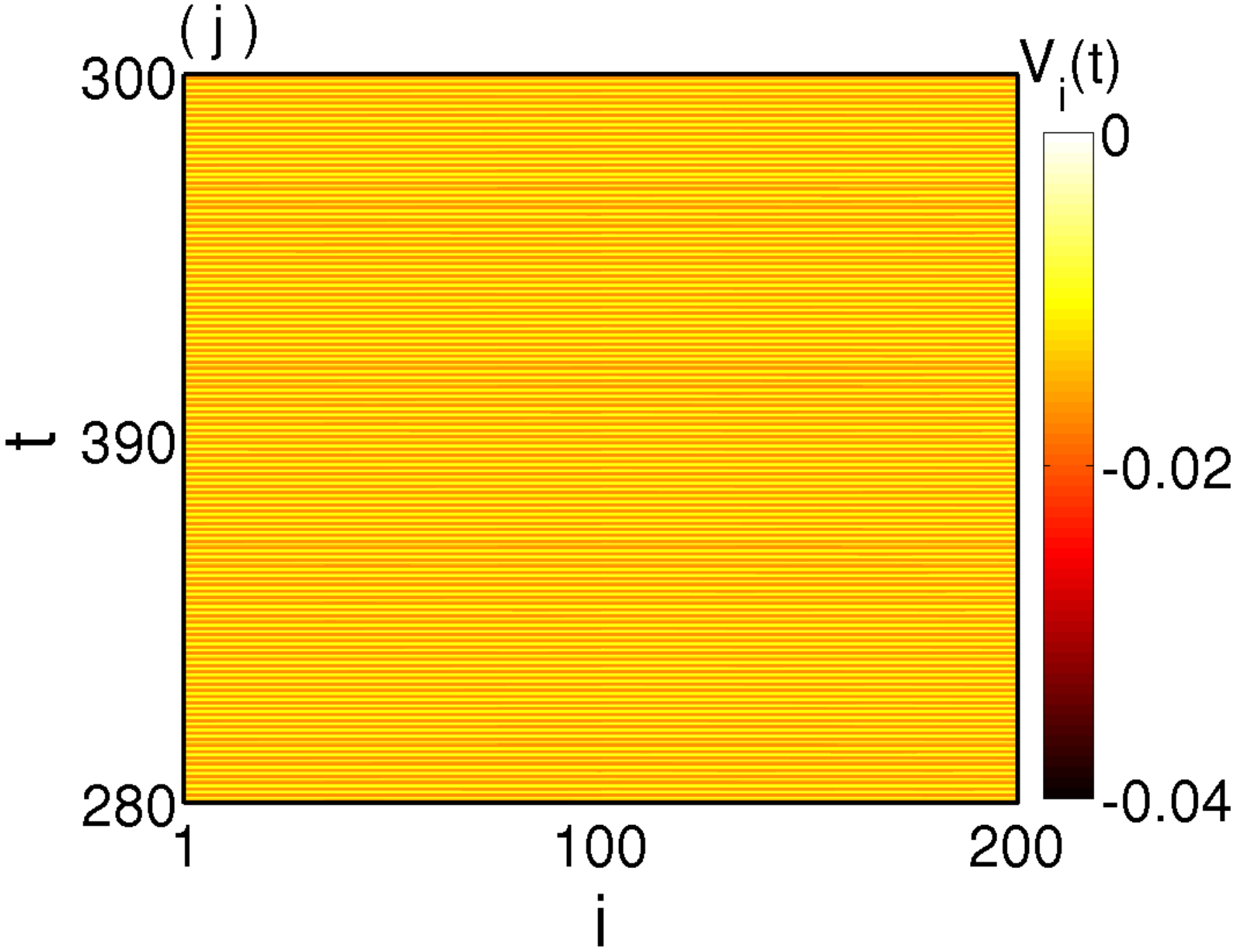}}
\caption{(Color online) Periodic initial conditions near complete synchronization manifold (left panel): (a) Variation of $V_{\mbox{SI}}$ (blue circle line), $\Lambda_{\mbox{GS}}$ (green square line), and $\Lambda_{\mbox{CS}}$ (red diamond line) with respect to $\epsilon$.  Snapshot of the state variable $V_i$ for (b) $\epsilon=15.0$ and (d) $\epsilon=27.0$ and corresponding spatio-temporal plots are in (c) and (e) respectively. PIs initial conditions (right panel): (f) Variation of $E_{\mbox{CS}}$ (left axis) and $B_{CS}$ (right axis) with respect to $\epsilon$. Snapshots at (g) $\epsilon=30.0$ and (i) $\epsilon=36.0$ and corresponding spatio-temporal behaviors are in (h) and (j) respectively. Insect figures show the corresponding trajectories in $(m_{K2_i}, V_i)$ plane.}
\label{figure_9}
\end{figure*}

\section{Stability of the complete synchronization states: master stability function approach}

\par Now, we investigate the persistence of the coherent states in terms of the auxiliary-system approach. For this purpose, we assume that the isolated evolution function $F({\bf x})$ satisfies the Lipschitz condition, i.e. there exists a positive constant $M$ such that
\begin{equation}
\begin{array}{l}
\label{eq_6}
\norm{F({\bf x})-F({\bf y})} \le M \norm{{\bf x}-{\bf y}}.
\end{array}
\end{equation}
The tenacity of the trivial equilibrium point of systems (\ref{eq_10}) implies the persistence of the coherent states. For this, we choose a Lyapunov function
\begin{equation}
\label{eq_7}
\begin{array}{lll}
G(t)=\frac{1}{2}\sum\limits_{i=1}^{N}{\bf z}_i^{tr}{\bf z}_i,
\end{array}
\end{equation}
where $tr$ denotes the transpose of a matrix. Taking the derivative of $G(t)$ with respect to time $t$ along the trajectory of the systems (\ref{eq_5}), we get
\begin{equation}
\begin{array}{lll}
\dot{G}(t)=\sum\limits_{i=1}^{N}{\bf z}_i^{tr}\dot{{\bf z}}_i\\\hspace{22pt}
=\sum\limits_{i=1}^{N}{\bf z}_i^{tr}[F({\bf x}_i+{\bf z}_i)-F({\bf x}_i)]-\epsilon \sum\limits_{i=1}^{N}{\bf z}_i^{tr}\Gamma {\bf z}_i
\end{array}
\end{equation}
Using the Lipschitz condition (\ref{eq_6}), it is easy to obtain the relation $[{\bf x}-{\bf y}]^{tr}[F({\bf x})-F({\bf y})] \le M [{\bf x}-{\bf y}]^{tr}[{\bf x}-{\bf y}]$. Then considering ${\bf Z}=[{\bf z}_1,{\bf z}_2,\cdots,{\bf z}_N]^{tr}$, we get\\
\begin{equation}
\begin{array}{lll}
\dot{G}(t)\le\sum\limits_{i=1}^{N}M{\bf z}_i^{tr}\dot{{\bf z}}_i-\epsilon \sum\limits_{i=1}^{N}{\bf z}_i^{tr}\Gamma {\bf z}_i=\big[M-\epsilon\lambda(I_N\otimes\Gamma)\big]{\bf Z}^{tr}{\bf Z},
\end{array}
\end{equation}
where $\otimes$ is the Kronecker matrix product and $\lambda(I_N\otimes\Gamma)$ denotes the smallest eigenvalue of the matrix $I_N\otimes\Gamma$. Now for the two square matrices $A$ and $B$ of order $N_1$ and $N_2$ respectively, if $\{\lambda_i:1\le i\le N_1\}$ and $\{\mu_j:1\le j\le N_2\}$ are the respective sets of eigenvalues, then the eigenvalues of $A\otimes B$ are $\lambda_i\mu_j$ for $i=1,2,\dots,N_1$ and $j=1,2,\dots,N_2$. This yields $\lambda(I_N\otimes\Gamma)=\lambda(I_N)\lambda(\Gamma)=\lambda(\Gamma)$. If all the eigenvalues of the inner coupling matrix $\Lambda$ are positive, then we get the persistence condition as
\begin{equation}
\begin{array}{lll}\label{eq_14}
\epsilon>\frac{M}{\lambda[\Gamma]}.
\end{array}
\end{equation}
By the Lasalle's invariance principle \cite{LaSalle}, the system (\ref{eq_10}) is asymptotically stable for these values of $\epsilon$, hence our original system oscillates coherently. The persistence condition (\ref{eq_14}) provides that once the coherent states are achieved, they will remain for any higher coupling strength if all the eigenvalues of the matrix $\Gamma$ are non-zero.

\par In our case, we observe that for higher values of $\epsilon$, qualitatively two different types of snapshots are exhibited: one is a smooth curve and the other is a straight line. The straight line snapshot indicates the complete synchronization among the coupled neurons. For this, we analytically derive the stability condition of the complete synchronization state, a special type of coherent state. For this type of coherent state, the snapshot of the oscillators exhibits a straight line. Each neuronal oscillator is identical and coupled by bidirectional electrical synapses. The individual evolution and interaction dynamics are both differentiable. Our coupling scheme is linear diffusive coupling; the complete synchronization manifold will be an invariant manifold. So the stability criterion of the complete synchronization states can be analyzed by the master stability function approach \cite{msf}, which gives the local stability criterion of the complete synchronization state. When complete synchronization occurs, let each node of the network evolve with the synchronization solution ${\bf x}_0(t)$, which satisfies the evolution equation $\dot{\bf x}=F({\bf x})$ from Eq. (\ref{eq_1}). To gauge the local stability of the complete synchronization state, we perturb the $i^{th}$ node from the complete synchronization state by an amount $\delta{\bf x}_i(t)$. Then the current state of the $i^{th}$ node is ${\bf x}_i={\bf x}_0+\delta{\bf x}_i$. Considering small perturbations and expanding around the complete synchronization solution up to first order, we get the linearized equation of the error systems for the complete synchronization state as
\begin{equation}
\begin{array}{l}
\label{eq_15}
\delta\dot{\bf x}_i=JF({\bf x}_0)\delta{\bf x}_i-\dfrac{\epsilon}{2P}\sum\limits_{j=1}^{N}\mathscr{L}_{ij}\Gamma\delta{\bf x}_j.
\end{array}
\end{equation}

\par To determine the transverse error components, we project the error vector $(\delta{\bf x}_1,\delta{\bf x}_2,\dots,\delta{\bf x}_N)$ to the space spanned by the Laplacian eigenvectors $U$. If $(\xi_1,\xi_2,\dots,\xi_N)^{tr}$ is the projected error vector, then $(\xi_1,\xi_2,\dots,\xi_N)^{tr}=\big(U\otimes I_d\big)^{-1}(\delta{\bf x}_1,\delta{\bf x}_2,\dots,\delta{\bf x}_N)^{tr}$, which yields the projected error dynamics as
\begin{equation}
\begin{array}{l}
\label{eq_16}
\dot{\xi}_i=[JF({\bf x}_0)-\frac{\epsilon}{2P}\gamma_i\Gamma]\xi_i,\hspace{10pt}i=1,2,\dots,N.
\end{array}
\end{equation}
The linearized equation for $i=1$, corresponds to the variational equation of the synchronization solution ${\bf x}_0$, which we have succeeded in separating from the other transverse directions. In our considered problem, the transverse error components can be written as
\begin{widetext}
\begin{equation}
\begin{array}{l}
\label{eq_17}
~~~\dot{\xi}_i^{V}=-\frac{1}{C}\Big[	 \bar{g}_{K2}m^2_{K2}+g_1+\bar{g}_{Na}\big(f(A_1,B_1,V)\big)^3 h_{Na}+3\bar{g}_{Na}\big(f(A_1,B_1,V)\big)^2f_V(A_1,B_1,V)(V-E_1)h_{Na}	 \Big]\xi_i^{V}\\[5pt]\hspace{35pt} -\frac{1}{C}\Big[2\bar{g}_{K2}m_{K2}(V-E_K )\xi_i^{m_{K2}}+\bar{g}_{Na}\big(f(A_1,B_1,V)\big)^3(V-E_{Na})\xi_i^{h_{Na}}	 \Big]-\dfrac{2\epsilon}{P}\sum\limits_{l=1}^{P}\sin^2\big(\frac{\pi l(i-1)}{N}\big)~\xi_i^{V},\\[8pt]
\dot{\xi}_i^{m_{K2}}=\dfrac{f_V(A_2,B_2+V^{shift}_{K2},V)\xi_i^{V}-\xi_i^{m_{K2}}}{\tau_{K2}},\\[8pt]
~\dot{\xi}_i^{h_{Na}}=\dfrac{f_V(A_3,B_3,V)\xi_i^{V}-\xi_i^{h_{Na}}}{\tau_{Na}},\hspace{50pt}i=2,3,\dots,N.
\end{array}
\end{equation}
\end{widetext}
Here $(V,m_{K2},h_{Na})$ satisfies the evolution equation of the synchronization solution (\ref{eq.1}). Next we investigate the effect of periodic initial conditions near and far away from the complete synchronization manifold by varying the coupling strength $\epsilon$.

\par Figure \ref{figure_9}(a) shows the variation of $V_{\mbox{SI}}$ (blue circle line), $\Lambda_{\mbox{GS}}$ (green square line), and $\Lambda_{\mbox{CS}}$ (red diamond line) for the range of $\epsilon\in[0,40]$ with PIs near the complete synchronization manifold. Initially, for a very small value of $\epsilon$, the network shows the incoherent state. After increasing the value of $\epsilon$, the transition from chimera to coherent states emerges at $\epsilon=8.0$. From here, the value of $V_{\mbox{SI}}$ becomes zero, and it remains unchanged for still higher values of $\epsilon$. This is in good agreement with our stability analysis of the coherent state, as $\Lambda_{\mbox{GS}}<0$ for $\epsilon\ge8.0$, while the values of $\Lambda_{\mbox{CS}}$ become negative at $\epsilon=25.0$. Hereafter, $\Lambda_{\mbox{CS}}$ maintains its negative values. So the complete synchronization state is locally stable for $\epsilon \ge 25.0$. To understand the qualitatively different coherent states, we plot the snapshots (middle panel) and corresponding spatio-temporal plots (lower panel) for two different values of $\epsilon$. At $\epsilon=15.0$, where $\Lambda_{\mbox{GS}}<0$ and $\Lambda_{\mbox{CS}}>0$, Fig. \ref{figure_9}(b) shows a smooth coherent profile, but it is not a straight line parallel to the $i-$axis. Therefore, at this coupling strength, the coupled systems are in coherent states, but not a complete synchronous state. The inset figure shows the phase-space in the $(m_{K2_i},V_i)$ plane. Here the network of coupled oscillators evolves with a quasi-periodic state, and Fig. \ref{figure_9}(c) shows the corresponding spatio-temporal plot. At another value of $\epsilon=27.0$, where $\Lambda_{\mbox{GS}}<0$ and $\Lambda_{\mbox{CS}}<0$, the snapshot in Fig. \ref{figure_9}(d) shows that the entire network is in a complete synchronization state. The complete synchronization solution is a period-1 limit cycle (inset figure), and the spatio-temporal plot of Fig. \ref{figure_9}(e) depicts its amplitude variations. The variation of $\Lambda_{\mbox{CS}}$ in the $(P, \epsilon)$ plane for a complete synchronization state is shown in the Appendix. For higher values of $\epsilon$, the complete synchronization state persists when the initial conditions are taken near the complete synchronization manifold.

\par However, this complete synchronization state is not globally stable for all $\epsilon$ in Fig. \ref{figure_9}(f) (left axis, blue circle curve). Particularly for the V-shaped PIs (far away from the complete synchronization manifold), the complete synchronization error
 \begin{widetext}
 	\begin{equation}
 	\begin{array}{l}
 	\label{eq_18}
 	E_{\mbox{CS}}=\Bigg\langle \sqrt{\dfrac{1}{N(N-1)}\sum\limits_{\substack{ i,j=1 \\ j\ne i }}^{N}\Big[(V_i-V_j)^2+(m_{K2_i}-m_{K2_j})^2+(h_{Na_i}-h_{Na_j})^2\Big]} \Bigg\rangle_t
 	\end{array}
 	\end{equation}
 \end{widetext}
first becomes zero at $\epsilon=25.0$. But it again shows a non-zero value for $\epsilon\in[29.0,33.0]$. In this range of $\epsilon$ with V-shaped initial conditions, $\Lambda_{\mbox{GS}}$ remains negative. Therefore, here the coherent state in the network is maintained. So here the complete synchronization state transforms to the generalized synchronization state. The complete synchronization state again returns at $\epsilon=33.0$ for these initial conditions.

\par Due to this multi-stable behavior of the complete synchronization state, we plot the basin stability $(B_{CS})$ of this state in Fig. \ref{figure_9}(f) (right axis, green square curve). The values of $B_{CS}$ remains zero up to $\epsilon=22.67$. By increasing $\epsilon$, the values of $B_{CS}$ gradually increases up to $\epsilon=26.67$, where it reaches the value $0.97$. Upon further increasing the values of $\epsilon$, it decreases up to $\epsilon=30.67$, where $B_{CS}=0.7578$. For higher values of $\epsilon$, the $B_{CS}$ again increases and finally reaches its maximum value $1.0$ at $\epsilon=34.67$. At $\epsilon=30.0$, however, $\Lambda_{\mbox{CS}}$ is negative in Fig. \ref{figure_9}(a) for small perturbations of their initial conditions. At $\epsilon=30.0$, the complete synchronization may not be stable (cf. Fig. \ref{figure_9}(f)), while the generalized synchronization state is stable (cf. Fig. \ref{figure_6}). Figures \ref{figure_9}(g) and \ref{figure_9}(h) show the snapshot and corresponding spatio-temporal plot for the generalized synchronization state. Again at $\epsilon=36.0$, the complete synchronization state is globally stable as $B_{CS}\backsimeq 1.0.$ Figure \ref{figure_9}(i) shows the snapshot for the complete synchronization state, where each neuron oscillates with the periodic solutions (inset figure), and their amplitude variations are shown in Fig. \ref{figure_9}(j).

\section{Conclusions}\label{conclusion}
In this paper, a neuronal network composed of Leech neurons assumed to be coupled non-locally through gap junctions has been studied. Leech neuronal oscillator exhibits the coexistence of periodic and chaotic attractors. These two stable states possess a well-separated basin of attraction in phase space. For numerical simulations, three types of `V' shaped initial conditions were chosen: a periodic basin, a chaotic basin, and a mixed basin. For a weaker synaptic strength, non-stationary chimera states emerge in this neuronal network. Three types of initial conditions possess different types of phase space of the entire network. Beyond a critical interaction strength, the chimera states become unstable, and coherent states appear, where the snapshot of the state variables follows a smooth profile. From this, we assert that in coherent states, the oscillators are in a generalized synchronization state. Using the auxiliary-system approach, we analytically derive the necessary condition for the chimera to coherent transitions, which are in excellent agreement with the value of the strength of incoherence. In a specified coupling range, the observed states are dependent on the initial conditions. By basin stability measurement, we have investigated the robustness of the chimera states with respect to the initial conditions. We also derived analytically the persistence conditions of the coherent states. This work contributes to a better understanding of the emergence of chimera states and their transition to coherent states in neuronal networks. Our analysis of the stability of coherent states will be useful in shedding light on the remarkable domain of chimera states.

\begin{acknowledgments}
DG was supported by DST-SERB (Department of Science and Technology), Government of India (Project no. EMR/2016/001039). MP was supported by the Slovenian Research Agency (Grants J4-9302, J1-9112, and P1-0403).
\end{acknowledgments}

\section{Appendix}

\begin{figure}
	\centerline{\includegraphics[scale=0.33]{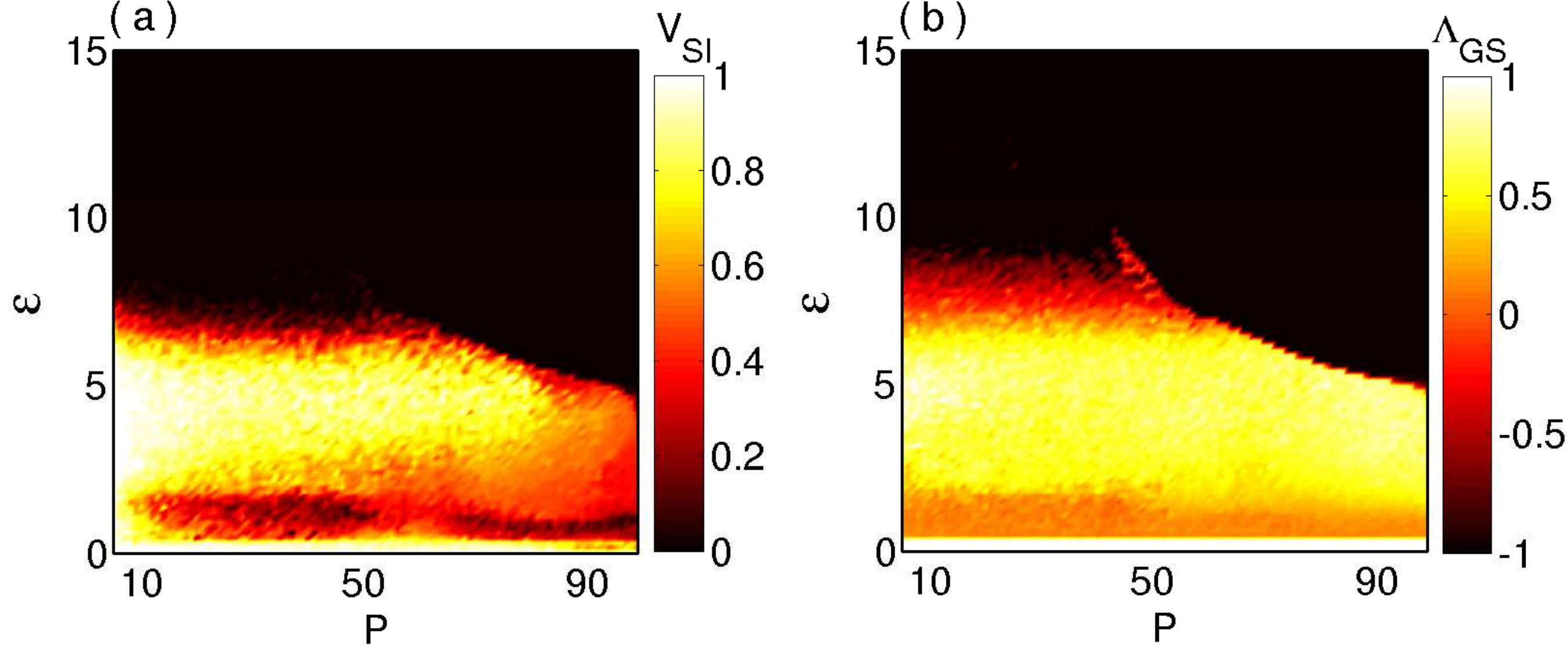}}
	\caption{(Color online) Two parameter phase diagram in $(P,\epsilon)$ plane with chaotic initial conditions. Color bars represent the variation of (a) $V_{\mbox{SI}}$ and (b) $\Lambda_{\mbox{GS}}$.}
	\label{figure_7}
\end{figure}
\begin{figure}[b]
	\centerline{\includegraphics[scale=0.33]{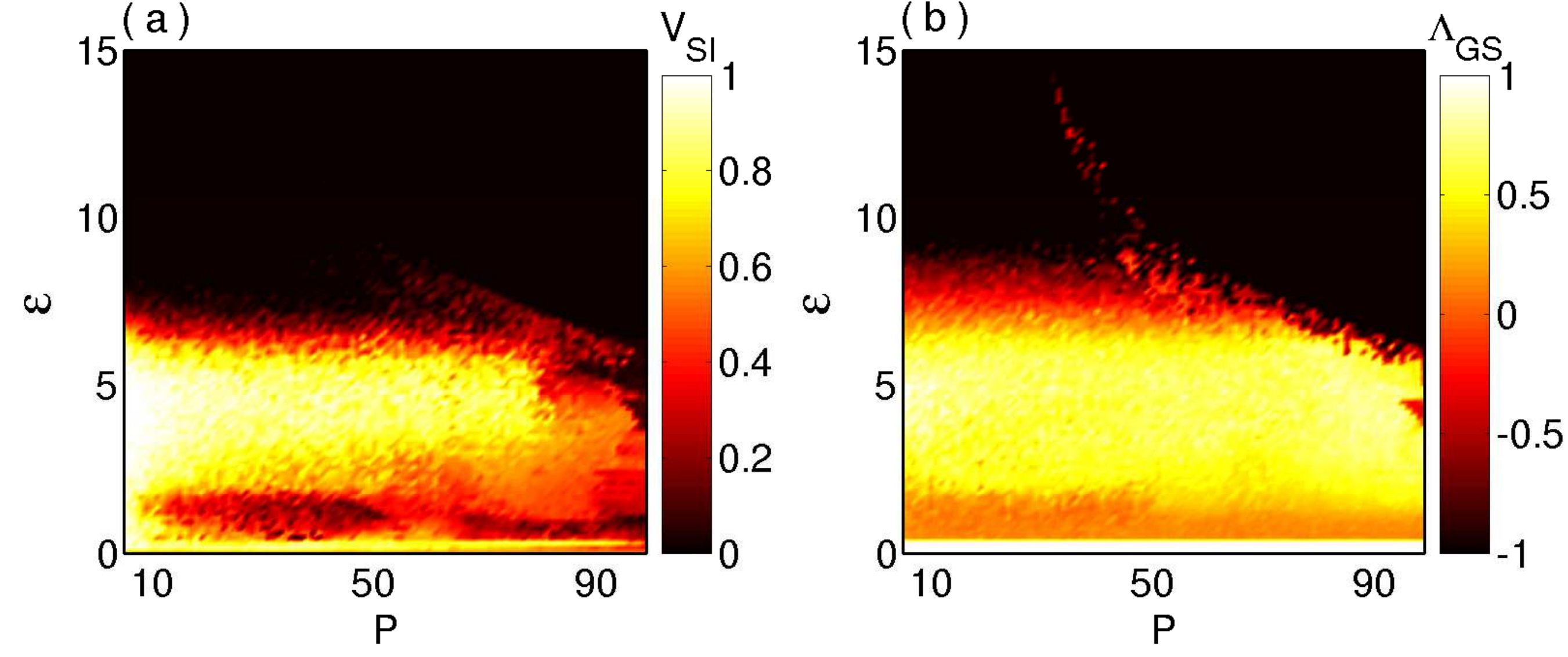}}
	\caption{(Color online) Two parameter phase diagram in $(P,\epsilon)$ plane with mixed initial conditions. Color bars represent the variation of (a) $V_{\mbox{SI}}$ and (b) $\Lambda_{\mbox{GS}}$.}
	\label{figure_8}
\end{figure}

In this appendix, we study the transition from chimera to coherent states in the $(P, \epsilon)$ plane using measurements of the strength of incoherence $V_{\mbox{SI}}$ and the maximum Lyapunov exponent for generalized synchronization $\Lambda_{\mbox{GS}}$. For this, we choose the initial conditions from the chaotic and mixed basins of attraction. Then we plot the $(P, \epsilon)$ plane for the complete synchronization state using the master stability function $\Lambda_{\mbox{CS}}$.

\par In the $(P,\epsilon)$ parameter space, the variations of $V_{\mbox{SI}}$ and $\Lambda_{\mbox{GS}}$ are drawn in Figs. \ref{figure_7}(a) and \ref{figure_7}(b), respectively, by taking CIs initial conditions. For lower values of $P$, the incoherent states persist up to $\epsilon\simeq5.5$, while for higher values of $P\ge10$, they persist up to $\epsilon=0.25$. Beyond that, chimera states emerge, that persist up to a certain value of the synaptic strength, after which coherent states emerge. The chimera to coherent transition point behaves non-monotonically up to $P=60$. After that, this critical value decreases monotonically as $P$ increases. For the incoherent and chimera regions, the values of $\Lambda_{\mbox{GS}}$ remain non-negative in Fig. \ref{figure_5}(b). For larger values of $\epsilon$, when coherent states appear the values of $\Lambda_{\mbox{GS}}$ becomes negative along with the $V_{\mbox{SI}}=0$.

\begin{figure}[b]
\centerline{\includegraphics[scale=0.45]{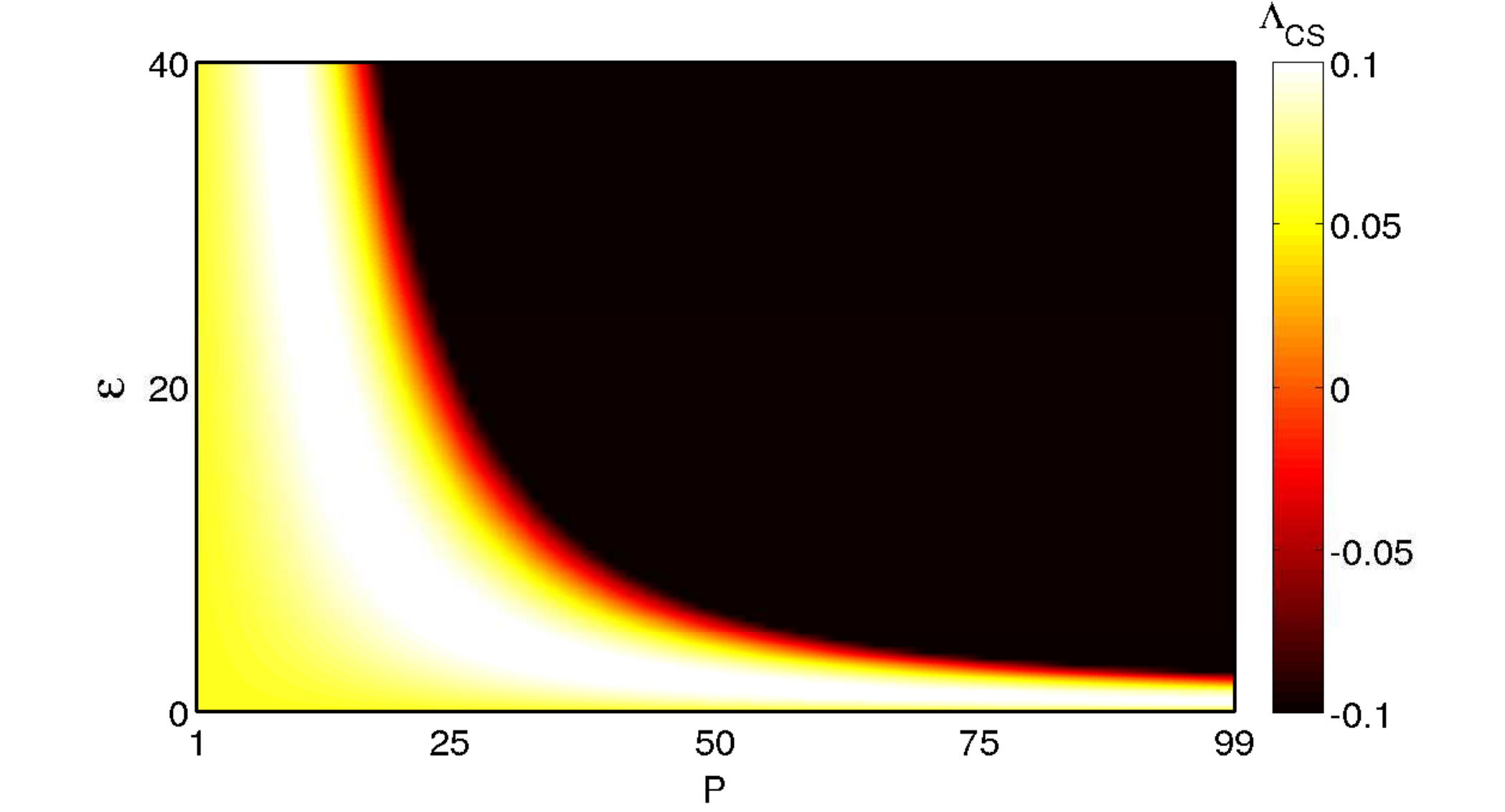}}
\caption{(Color online) Variation of the master stability function $\Lambda_{\mbox{CS}}$ for complete synchronization state in $(P,\epsilon)$ parameter plane for periodic initial conditions.}
\label{figure_10}
\end{figure}

\par Figures \ref{figure_8}(a) and \ref{figure_8}(b) respectively represent the values of $V_{\mbox{SI}}$ and $\Lambda_{\mbox{GS}}$ in the $(P,\epsilon)$ plane by taking MIs-type initial conditions. Both subplots are in excellent agreement with each other. For lower values of $P$, the incoherent states persist up to $\epsilon\simeq5.5$, but for higher values of $P$, chimera states appear once we switch on the interaction. When Fig. \ref{figure_8}(a) shows incoherent or chimera states, the value of $\Lambda_{\mbox{GS}}$ is positive there. For higher values of $\epsilon$, as soon as chimera states annihilate and coherent states emerge, the values of $\Lambda_{\mbox{GS}}$ become negative.

Figure \ref{figure_10} represents the parameter region in the $(P,\epsilon)$ plane for the complete synchronization states, by taking PIs initial conditions. The color bar shows the variation of the master stability function $\Lambda_{\mbox{CS}}$. The complete synchronization states are enhanced monotonically with respect to both the varying parameter $p$ and $\epsilon$.


\begin{thebibliography}{53}%
\bibitem{r12}Y. Kuramoto and D. Battogtokh, Nonl. Phen. Compl. Syst. {\bf 5}, 380 (2002).
\bibitem{r13}D. M. Abrams and S. H. Strogatz, Phys. Rev. Lett. {\bf 93}, 174102 (2004).
\bibitem{r15a}S. Majhi, B. K. Bera, D. Ghosh, and M. Perc, Phys. Life Rev. {\bf 28}, 100 (2019).
\bibitem{chimera_r1} M. J. Panaggio, and D. M. Abrams, Nonlinearity {\bf 28}, R67 (2015).
\bibitem{r14a}B. K. Bera, S. Majhi, D. Ghosh, and M. Perc, Europhys. Lett. {\bf 118}, 10001 (2017).
\bibitem{chimera_r2} O E Omel’chenko, Nonlinearity {\bf 31}, R121 (2018).
\bibitem{r18}N. S. Frolov, V. A. Maksimenko, V. V. Makarov, D. V. Kirsanov, A. E. Hramov, and J. Kurths, Phys. Rev. E {\bf 98}, 022320 (2018).
\bibitem{r19}S. Ghosh, A. Kumar, A. Zakharova, and S. Jalan, Europhys. Lett. {\bf 115}, 60005 (2016).
\bibitem{r20}T. Kapitaniak, P. Kuzma, J. Wojewoda, K. Czolczynski, and Y. Maistrenko, Sci. Rep. {\bf 4}, 6379 (2014).
\bibitem{r21}E. A. Martens, S. Thutupalli, A. Fourri`ere, and O. Hallatschek, Proc. Natl. Acad. Sci. USA {\bf 110}, 10563 (2013).
\bibitem{r22}S. Nkomo, M. R. Tinsley, and K. Showalter, Phys. Rev. Lett. {\bf 110}, 244102 (2013).
\bibitem{amplt_cmra} A. Zakharova, M. Kapeller, and E. Sch\"{o}ll, Phys. Rev. Lett. {\bf 112}, 154101 (2014).
\bibitem{r31}E. A. Martens, C. R. Laing, and S. H. Strogatz, Phys. Rev. Lett. {\bf 104}, 044101 (2010).
\bibitem{epjst} S. Kundu, S. Majhi, P. Muruganandam, and D. Ghosh, Eur. Phys. J. Spec. Top. {\bf 227}, 983 (2018).
\bibitem{r25}A. Zakharova, M. Kapeller, and E. Sch{\"o}ll, J. Phys. Conf. Ser. {\bf 727}, 012018 (2016).
\bibitem{r40}S. Kundu, B. K. Bera, D. Ghosh, and M. Lakshmanan, Phys. Rev. E {\bf 99}, 022204 (2019).
\bibitem{r46}B. K. Bera, D. Ghosh, and T. Banerjee, Phys. Rev. E {\bf 94}, 012215 (2016).
\bibitem{r29}J. Hizanidis, E. Panagakou, I. Omelchenko, E. Sch{\"o}ll, P. H{\"o}vel, and A. Provata, Phys. Rev. E {\bf 92}, 012915 (2015).
\bibitem{r30}S. Majhi and D. Ghosh, Chaos {\bf 28}, 083113 (2018).
\bibitem{spike} B. K. Bera, S. Rakshit, D. Ghosh, and J. Kurths, Chaos {\bf 29}, 053115 (2019).
\bibitem{chimera_global} G. C. Sethia, A. Sen, and G. L. Johnston, Phys. Rev. E {\bf 88}, 042917 (2013).
\bibitem{r39}B. K. Bera, D. Ghosh, and M. Lakshmanan, Phys. Rev. E {\bf 93}, 012205 (2016).
\bibitem{cmra_eco1}P. S. Dutta, and T. Banerjee, Phys. Rev. E {\bf 92}, 042919 (2015).
\bibitem{cmra_SQUID}N. Lazarides, G. Neofotistos, and G. P. Tsironis, Phys. Rev. B {\bf 91}, 054303 (2015).
\bibitem{cmra_quantum}V. M. Bastidas, I. Omelchenko, A. Zakharova, E. Schöll, and T. Brandes, Phys. Rev. E {\bf 92}, 062924 (2015).
\bibitem{r7}G. Cymbalyuk and A. Shilnikov, J. Comput. Neurosci. {\bf 18}, 255 (2005).
\bibitem{r9}C. Canavier, D. Baxter, J. Clark, and J. Byrne, J. Neurophysiol. {\bf 69}, 2252 (1993).
\bibitem{r8}C. M. Gray and W. Singer, Proc. Natl. Acad. Sci. USA {\bf 86}, 1698 (1989).
\bibitem{r1}E. Yilmaz, M. Ozer, V. Baysal, and M. Perc, Sci. Rep. {\bf 6}, 30914 (2016).
\bibitem{r2} P. Dayan, {\it Theoretical Neuroscience- Computational and Mathematical Modeling of Neural Systems}, MIT Press (2005).
\bibitem{r6}M. Gosak, R. Markovi{\v{c}}, J. Dolen{\v{s}}ek, M. S. Rupnik, M. Marhl, A. Sto{\v{z}}er, and M. Perc, Phys. Life Rev. {\bf 24}, 118 (2018).
\bibitem{r10}H. Lechner, D. Baxter, J. Clark, and J. Byrne, J. Neurophysiol. {\bf 75}, 957 (1996).
\bibitem{r32}S. Majhi, M. Perc, and D. Ghosh, Sci. Rep. {\bf 6}, 39033 (2016).
\bibitem{r33} R. G. Andrzejak, C. Rummel, F. Mormann, and K. Schindler, Sci. Rep. {\bf 6}, 23000 (2016).
\bibitem{r34} R. Gopal, V. K. Chandrasekar,  A. Venkatesan, and M. Lakshmanan, Phys. Rev. E {\bf 89}, 052914 (2014).
\bibitem{r35}D. P. Rosin, D. Rontani, N. D. Haynes, E. Sch{\"o}ll, and D. J. Gauthier, Phys. Rev. E 90, 030902 (2014).
\bibitem{cmra_unih1} N. C. Rattenborg, C. J. Amlaner, and S. L. Lima, Neurosci. Biobehav. Rev. {\bf 24}, 817 (2000).
\bibitem{cmra_unih2} N. C. Rattenborg, Naturwissenschaften {\bf 93}, 413 (2006).
\bibitem{r37}A. V{\"u}llings, J. Hizanidis, I. Omelchenko, and H{\"o}vel, New J. Phys. {\bf 16}, 123039 (2014).
\bibitem{r37a}S. Kundu, S. Majhi, B. K. Bera, D. Ghosh, and M. Lakshmanan, Phys. Rev. E {\bf 97}, 022201 (2018).
\bibitem{r38}I. Omelchenko, E. Omel\textquotesingle~chenko, P. H{\"o}vel, and E. Sch{\"o}ll, Phys. Rev. Lett. {\bf 110}, 224101 (2013).
\bibitem{r38a}N. Semenova, A. Zakharova, V. Anishchenko, and E. Sch{\"o}ll, Phys. Rev. Lett. {\bf 117}, 014102 (2016).
\bibitem{neuronal_nl} B. Ermentrout, J. Campbell, and G. Oster, Veliger {\bf 28}, 369 (1986).
\bibitem{ocular_nl} N. V. Swindale, Proc. R. Soc. Lond. B {\bf 208}, 243 (1980).
\bibitem{jj_nl} J. R. Phillips, H. S. J. van der Zant, J. White, and T. P. Orlando, Phys. Rev. B {\bf 47}, 5219 (1993).
\bibitem{r41}Z. Wei, F. Parastesh, H. Azarnoush, S. Jafari, D. Ghosh, M. Perc, and M. Slavinec, Europhys. Lett. {\bf 123}, 48003 (2018).
\bibitem{ausys_3} N. F. Rulkov, M. M. Sushchik, L. S. Tsimring, and H. D. I. Abarbanel, Phys. Rev. E {\bf 51}, 980 (1995).
\bibitem{ausys_1} H. D. I. Abarbanel, N. F. Rulkov, and M. M. Sushchik, Phys. Rev. E {\bf 53}, 4528 (1996).
\bibitem{ausys_4} L. Kocarev and U. Parlitz, Phys. Rev. Lett. {\bf 76}, 1816 (1996).
\bibitem{ausys_2} Z. Zheng, X. Wang, and M. C. Cross, Phys. Rev. E {\bf 65}, 056211 (2002).
\bibitem{hung}Y. C. Hung, Y. T. Huang, M. C. Ho, and C. K. Hu, Phys. Rev. E {\bf 77}, 016202 (2008).
\bibitem{guan}S. Guan, X. Wang, X. Gong, K. Li, and C. H. Lai, Chaos {\bf 19}, 013130 (2009).
\bibitem{chen}J. Chen, J. Lu, X. Wu, and W. X. Zheng, Chaos {\bf 19}, 043119 (2009).
\bibitem{liu}H. Liu, J. Chen, J. Lu, and M. Cao, Physica A {\bf 389}, 1759 (2010).
\bibitem{gs_njp}S. Guan, X. Gong, K. Li, Z. Liu, and C. H. Lai, New J. Phys. {\bf 12}, 073045 (2010).
\bibitem{olga} O. I. Moskalenko, A. A. Koronovskii, and A. E. Hramov, Phys. Rev. E {\bf 87}, 064901 (2013).
\bibitem{cmra_bs1} E. A. Martens, M. J. Panaggio, and D. M. Abrams, New J. Phys. {\bf 18}, 022002 (2016).
\bibitem{cmra_srep} S. Rakshit, B. K. Bera, M. Perc and D. Ghosh, Sci. Rep. {\bf 7}, 2412 (2017).
\bibitem{bs_delay} S. Leng, W. Lin and J. Kurths, Sci. Rep. {\bf 6}, 21449 (2016).
\bibitem{LaSalle}H. K. Khalil, {\it Nonlinear Systems}, (New Jersey: Prentice-Hall) (2002).
\bibitem{msf}L. M. Pecora and T. L. Carroll, Phys. Rev. Lett. {\bf 80}, 2109 (1998).
\end{thebibliography}
\end{document}